\documentclass[%
 reprint,
 amsmath,amssymb,
 aps,
]{revtex4-1}

\usepackage{graphicx}
\usepackage{dcolumn}
\usepackage{bm}
\usepackage{lipsum}

\usepackage{subfigure}

\begin{document}


\title{
Effect of inertia on the evasion and pursuit dynamics of prey swarms and the emergence of an optimal mass ratio for the predator-prey arms race\\
}

\author{Dipanjan Chakraborty, Arkayan Laha, and Rumi De}
\email{Corresponding author:rumi.de@iiserkol.ac.in}
\affiliation{%
Department of Physical Sciences, Indian Institute of Science Education and Research Kolkata,
 Mohanpur - 741246, Nadia, West Bengal, India\\
 }


\begin{abstract} 

We show, based on a theoretical model, how inertia plays a pivotal role in the survival dynamics of a prey swarm while chased by a predator. With the varying mass of the prey and predator, diverse escape patterns emerge, such as circling, chasing, maneuvering, dividing into subgroups, and merging into a unitary group, similar to the escape trajectories observed in nature. Moreover,  we find a transition from non-survival to survival of the prey swarm with increasing predator mass. The transition regime is also sensitive to the variation in prey mass. Further, the analysis of the prey group survival as a function of predator-to-prey mass ratio unveils the existence of three distinct regimes: (i) frequent chase and capture leading to the non-survival of the prey swarm, (ii) an intermediate regime where competition between pursuit and capture occurs, resembling an arms race, and (iii) the survival regime without the capture of prey. Interestingly, our study demonstrates the existence of a favourable predator-prey mass ratio for efficient predation, which corroborates with the field studies.

\end{abstract}

\pacs{Valid PACS appear here}
\maketitle


\section*{Introduction}

Swarming of insects, flocking of birds, schooling of fishes, and migration of coordinated cells are examples of various collective behaviors observed in nature \cite{ Kparrishscience, krausebook2002, vijay2020, sumpterbook2010, ballerini2008, Partho2019, Arnoldplosone2012, jmichaelscience2009,  debangana2019, debangana2022}. A variety of reasons have been suggested behind the origin of collective animal motion, such as searching for food, finding new habitat, mating and reproduction, etc \cite{Arnoldplosone2012, traniello2003ARE, Dipanjan-Review2022}. 
Another important reason is the defensive strategy often opted by the prey to stay in a group to protect themselves from the predator 
attacks. Moving in a group enhances the chances of prey survival as many can keep a watch on possible attacks; thus, the overall vigilance increases \cite{neilletal, Kparrishscience, penzhornbook1984, pitcherbook1983}. However, cohesive motion sometimes can make the predation easier as the prey group could be quickly spotted by the predator compared to an individual prey \cite{parrishenvbio}. 
Many studies have been conducted to get insights into the complex prey-predator dynamics and the intrinsic interactions among individuals. In terms of theoretical studies, 
the lattice models remain one of the popular prey-predator model frameworks that have provided many insights into 
the pursuit and evasion of the predator and the prey \cite{Oshaninproceedings2009, Siddharth2020}.
Besides, many off-lattice models, following the Vicsek model on flocking  \cite{vicsekprl1995},  have also been developed to understand the prey-predator interactions. Based on the self-propelled particle models, incorporating short-range repulsion and long-range attractions among the prey group and intergroup interactions between the prey-predator, many aspects of the predation and efficient escape strategies have been studied \cite{AngelaniPRL2012, Lettetal2014theoeco, Chen2014}. Further, a recent study by our group has shown that the range of cooperative interaction within the prey swarm plays a crucial role in determining the survival chances of the prey group and the emergence of various escape trajectories \cite{chakraborty2020}.

However, most previous theoretical models have generally ignored the influence of inertial forces on the prey-predator dynamics.
A recent experimental study on two predator-prey pairs, lion-zebra and cheetah-impala, in their natural savannah habitat, demonstrates that the predators have more muscle power and a higher ability to accelerate and decelerate than their respective prey \cite{wilson2018nature}. On the other hand, prey are able to escape using their turning maneuverability, which depends on the inertial forces. This way, it allows the scope of survival of both the prey and the predator species, exhibiting an evolutionary arms race. Besides, many field studies suggest the existence of an optimal mass ratio of the predator and the prey in natural ecosystems that signifies inertia plays a crucial role in setting up the food chain in an ecosystem  \cite{barnes2010ecology, woodward2007bookchapter, brose2010funeco, brose2008janimalecology}.
There are also other studies that demonstrate the influence of inertia on various other dynamical systems. For example, it has been shown that inertia is essential in the global ordering and stability of flocks of microorganisms in a turbulent environment \cite{choudhary2015epl}. Scholz {\it et al.} have found in experimental and theoretical observations that the inertia causes
a distinct delay between the orientation and velocity of macroscopic self-propelled vibrobot particles \cite{Scholzetal2018natcomm}.
Further, standard models on flocking, such as Vicsek model and its different variants, could not describe the collective turning of flocks as the inertial contribution is ignored \cite{vicsekprl1995}. Later, Cavagna {\it et al.} have developed a model including the inertial forces that successfully captured the collective turning and information propagation as observed in natural flocks \cite{cavagnaetal2015jsp}.

Until now, only a very few models have investigated the influence of inertia on predator-prey dynamics. Zhdankin and Sprott have simulated the prey-predator behaviours considering them as agents which interact via radial force laws \cite{ZhdankinPRE2010}. They have observed periodic, quasiperiodic, and chaotic solutions with varying friction and the mass of the prey-predator system. Another agent-based model, considering the inertial forces, has shown how a group of predators could strategize their chase to catch a faster prey within a closed, soft boundary, and there is an optimal group size for efficient capture \cite{milanjanosov2017njp}.
However, these models did not look into the complex swarming dynamics of prey groups under the predator attack.
Prey swarm shows various emergent escape trajectories
such as circling, F-maneuvering, splitting up into subgroups, and merging again into one group as a part of their survival strategy \cite{kerleyjoz2005, Mckenzieinterfacefocus2012, carobook2005, Humpherisoecologia1970, DomeniciJEB2011, EdutBBR2004, DomeniciMB1997}.
In this paper, we investigate how inertia influences the survival dynamics and the emergence of diverse escape trajectories of a prey swarm when chased by a predator. We consider the framework of a self-propelled particle model and incorporate the prey-prey and prey-predator interactions by the pairwise attractive and repulsive forces.
First, we study the effect of inertial forces on the steady-state in the case of a weak predator where the predator is mostly unable to capture the prey. It exhibits a transition from stable ring formation by the prey swarm around the predator to chasing dynamics with varying mass. These numerical findings are also substantiated by stability analysis calculations. 
Further, we explore the evasion and pursuit dynamics of a prey swarm under the attack of a strong predator. Diverse escape trajectories emerge as we vary the mass of the prey and the predator. 
Moreover, it shows a transition from non-survival to survival of the prey group as the mass of the predator increases. We have also investigated how the predator-to-prey mass ratio affects these transition regimes. Interestingly, our analysis shows a favourable predator-to-prey mass ratio for the competitive chasing dynamics where some of the prey survive and some are killed.
We further probe how the
variation of predator strength and the prey group size affects the optimal value of the predator-to-prey mass ratio for efficient predation.\\
 
\section*{Theoretical Model}
  \begin{figure}[!t]
    	\centering
    	\includegraphics[width=7.5cm,height=6.5cm]{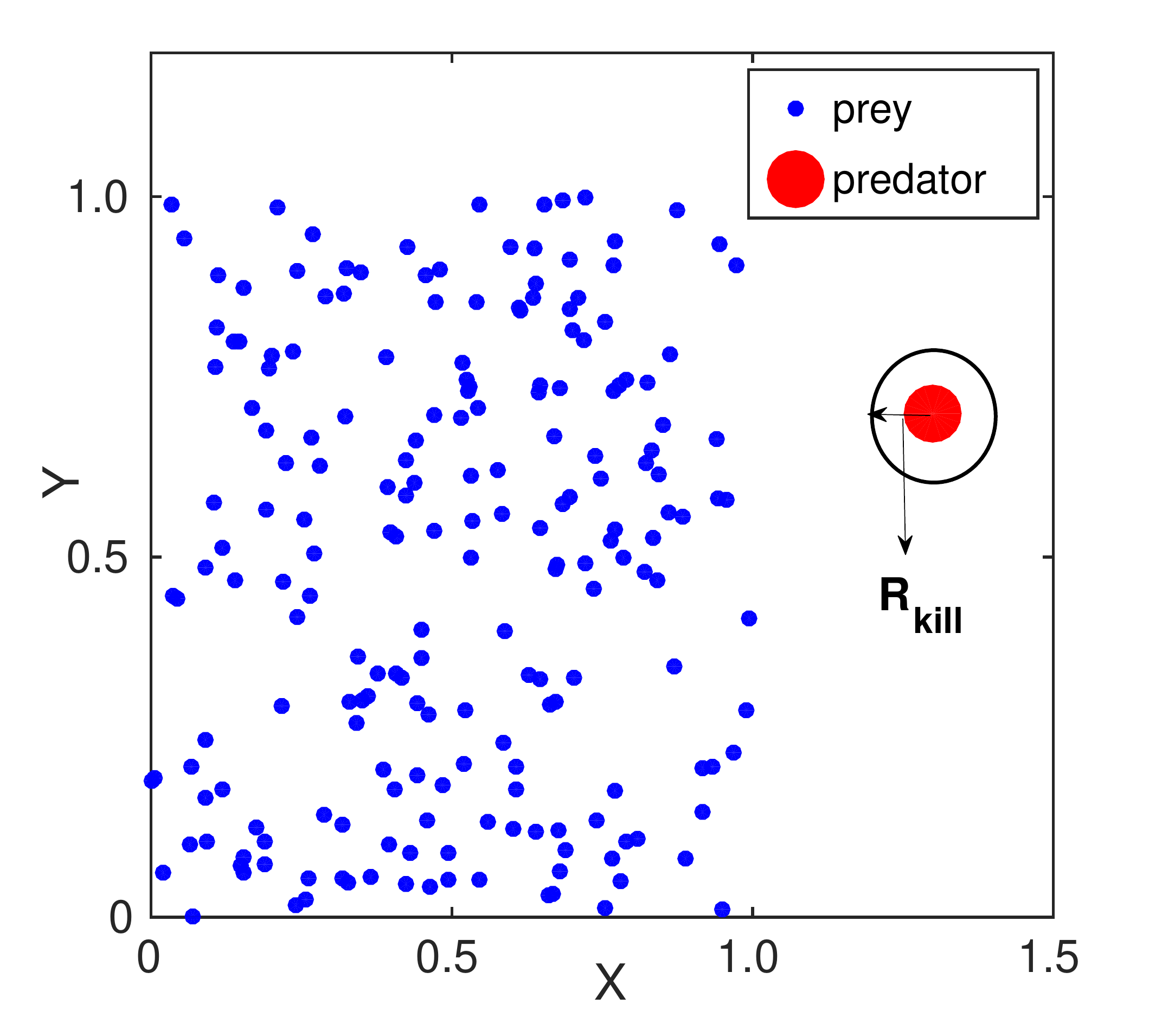}
    	\caption{ Initially, N number of prey have been randomly positioned in a unit square as shown in solid blue dots. The predator starts chasing the prey group from a nearby position, as shown by a big red circle. Kill radius has been shown by $R_{\rm kill}$,  surrounding the predator (color online).}
    	\label{fig:Initialconfig}
    \end{figure} 

In our model, each prey and predator are considered as active particles. Initially,  N number of prey are positioned randomly in a unit square in two-dimensional space, and the predator starts chasing the prey group from a nearby place as shown in Fig.  (\ref{fig:Initialconfig}). The escape and chasing dynamics depend on the interactions within the prey swarm and between the prey and the predator.
The equation of motion of $i$'th prey is given as follows,
\begin{equation} 
m_{\rm pr}\frac{d^2\vec{r}_i}{dt^2}=-\mu_{\rm pr}\frac{d\vec{r}_i}{dt} + \vec{F}_{i,\rm prey-prey} + \vec{F}_{i,\rm prey-predator}. 
\label{ini_preyequation}
\end{equation}
Here, $\vec{r}_i$ denotes the position of the $i$'th prey, and $m_{\rm pr}$ represents the mass of each prey.
As the prey moves, it experiences a viscous force arising from the environment, where $\mu_{\rm pr}$ is the coefficient of such frictional force.
  Apart from the frictional force term, the other two terms in the right-hand side of the Eq. (\ref{ini_preyequation}) indicate the interaction forces between the prey-prey and the prey-predator. 
We have modeled the prey-prey interaction by pairwise short-range repulsion and long-range attraction forces.
Short-range repulsive force helps the prey group to avoid collisions among themselves, and long-range attractive force keeps the prey group cohesive. Thus, the prey-prey interaction force of the $i$'th prey, $\vec{F}_{i,\rm prey-prey}$,  is given by averaging over all interacting prey within the prey swarm as, 
\begin{equation}
 	\vec{F}_{i,\rm prey-prey}=\frac{1}{N_{\rm sur}}\sum_{j=1}^{N_{\rm sur}}({\alpha \frac{\vec{r}_i-\vec{r}_j}{|\vec{r}_i -\vec{r}_j|^2} - \beta (\vec{r}_i - \vec{r}_j)}),
\end{equation} 
where $\alpha$ represents the strength of short-range repulsion, $\beta$ is the strength of long-range prey-prey attraction, and $N_{\rm sur}$ is the total number of survived prey in the prey group.
Since it is the instinct of each prey to move away from the predator to survive, we have modeled the prey-predator interaction, $\vec{F}_{i,\rm prey-predator}$, by a repulsive radial force as, 
\begin{equation}
\vec{F}_{i,\rm prey-predator}=\gamma \frac{\vec{r}_i - \vec{r}_p}{|\vec{r}_i - \vec{r}_p|^2}.
\end{equation}
Here, $\vec{r}_p$ is the position of the predator, and $\gamma$ signifies the strength of the repulsive force between the prey and predator.

Similarly, the equation motion of the predator can be written as,
\begin{equation}
m_{\rm pd}\frac{d^2\vec{r}_p}{dt^2}=-\mu_{\rm pd}\frac{d\vec{r}_p}{dt} + \vec{F}_{\rm predator-prey};
\label{ini_predatorequation}
\end{equation}
where $m_{\rm pd}$ denotes the mass of the predator and $\mu_{\rm pd}$ is the strength of the frictional force acting on the predator.
$\vec{F}_{\rm predator-prey}$ represents the predator-prey attraction force.
Since the predator chases the prey swarm from a close proximity, it could see all prey and strategize the move based on the interactions averaged over all survived prey which is modeled as, 
\begin{equation}
\vec{F}_{\rm predator-prey}= - \frac{\delta}{N_{\rm sur}} \sum_{i=1}^{N_{\rm sur}} \frac{\vec{r}_p - \vec{r}_i}{|\vec{r}_p - \vec{r}_i|^3}.
\end{equation}
Here, the strength of the predator is represented by $\delta$. We have introduced a kill radius of the predator to incorporate the capture of prey by the predator. When prey comes within the circular area covered by the predator's kill radius [as illustrated in Fig. (\ref{fig:Initialconfig})], it is killed. Moreover, the predator-prey attraction force dominates over the prey-predator repulsion when the predator is closer to the prey (due to the difference in exponents in the denominator of the interaction terms). If the predator starts far away from the prey group, prey-predator repulsive strength overpowers the prey-predator attraction, so the prey group outruns the predator and escapes. Hence, competition between the attractive and repulsive forces among the prey and the predator leads to a survival race. 
Along with the interaction forces, the inertial force also plays a crucial role in the turning and maneuverability of the prey and predator, hence determining the survival outcome.

To simplify the analysis, we study the dynamics in dimensionless units, and the scaled equations are given as follows,
\begin{equation}
\begin{split}
M_{\rm pr}\frac{d^2\vec{R}_i}{dT^2}= & -\frac{d\vec{R}_i}{dT} + \frac{1}{N_{\rm sur}}\sum_{j=1}^{N_{\rm sur}}({\alpha_{0} \frac{\vec{R}_i-\vec{R}_j}{|\vec{R}_i -\vec{R}_j|^2} - \beta_{0} (\vec{R}_i - \vec{R}_j)}) \\ 
 & + \gamma_{0} \frac{\vec{R}_i - \vec{R}_p}{|\vec{R}_i - \vec{R}_p|^2},
\end{split}
\label{preyequation}
\end{equation}
\begin{equation}
M_{\rm pd}\frac{d^2\vec{R}_p}{dT^2}=-\frac{d\vec{R}_p}{dT} -\frac{\delta_0}{N_{\rm sur}} \sum_{i=1}^{N_{\rm sur}} \frac{\vec{R}_p - \vec{R}_i}{|\vec{R}_p - \vec{R}_i|^3}.
\label{predequation}
\end{equation}

Here, $\vec{R}_i=\vec{r}_i/l_0$, $\vec{R}_j=\vec{r}_j/l_0$,  $\vec{R}_p=\vec{r}_p/l_0$, and $T=t/\tau$ 
are dimensionless position and time variables; $l_{0}$ and $\tau$ represent the characteristic length scale and time scale of the system.   
$M_{\rm pr}=m_{\rm pr}/\tau \mu_{\rm pr}$ and  $M_{\rm pd}=m_{\rm pd}/\tau \mu_{\rm pd}$ denote the scaled mass of the prey and the predator respectively. 
From now, by \lq mass' we refer to the scaled mass in the rest of the paper. Also, the other scaled parameters are given as, $\alpha_0=\alpha \tau/l_0^2 \mu_{\rm pr}$, $\beta_0=\beta \tau/\mu_{\rm pr}$, $\gamma_0=\gamma \tau/\mu_{\rm pr} l_0^2$, and $\delta_0=\delta \tau/\mu_{\rm pd} l_0^3$.

\section*{Results}

\begin{figure*}[!t]
 	\subfigure[ ]{\label{fig:mpr_1pt0_mpd_2pt0_t400}\includegraphics[width=3.5cm,height=3.5cm]{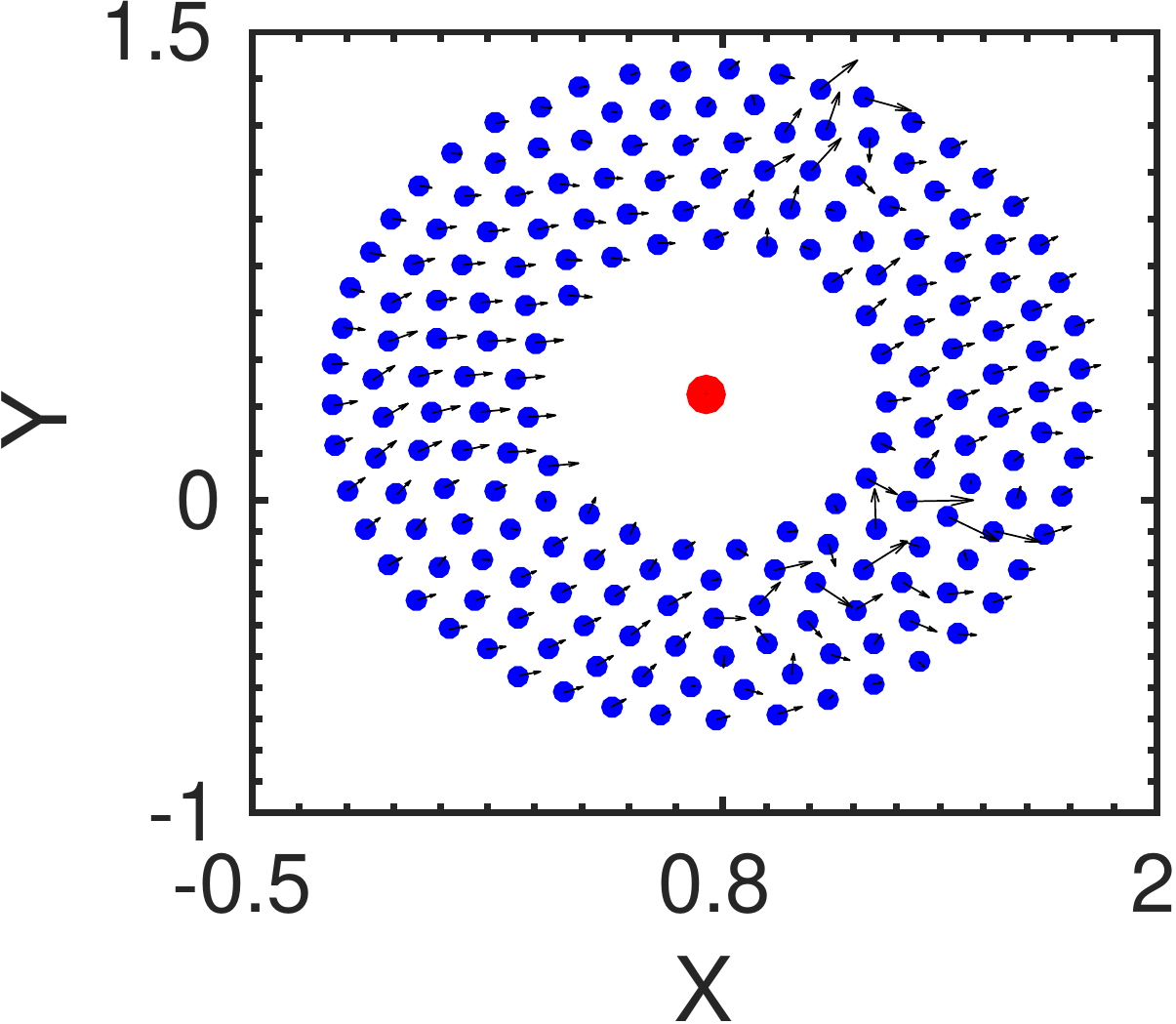}}
 	\subfigure[ ]{\label{fig:mpr_1pt0_mpd_3pt0_t300}\includegraphics[width=3.5cm,height=3.5cm]{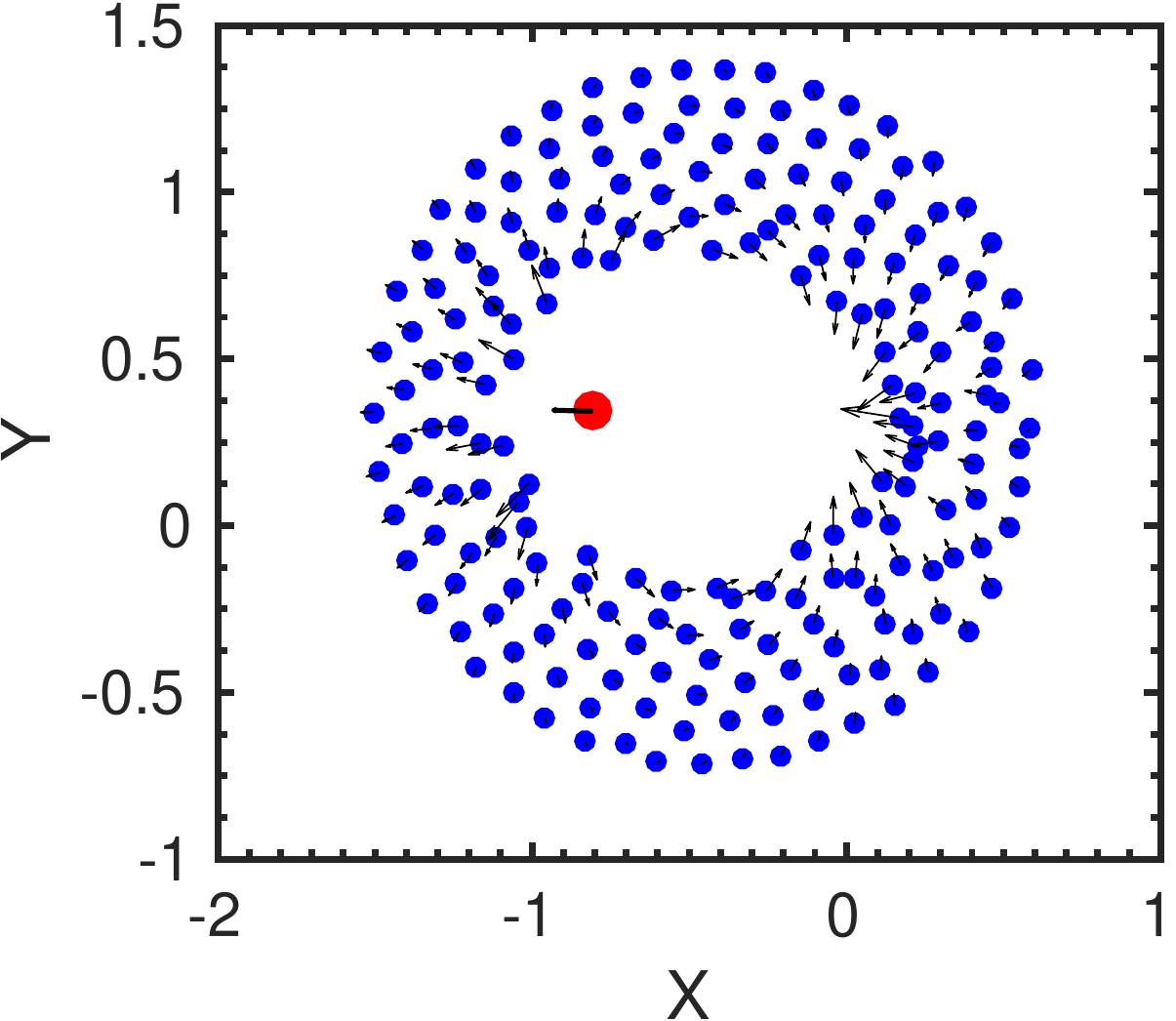}}
  	\subfigure[ ]{\label{fig:mpr_1pt0_mpd_3pt0_t310}\includegraphics[width=3.5cm,height=3.5cm]{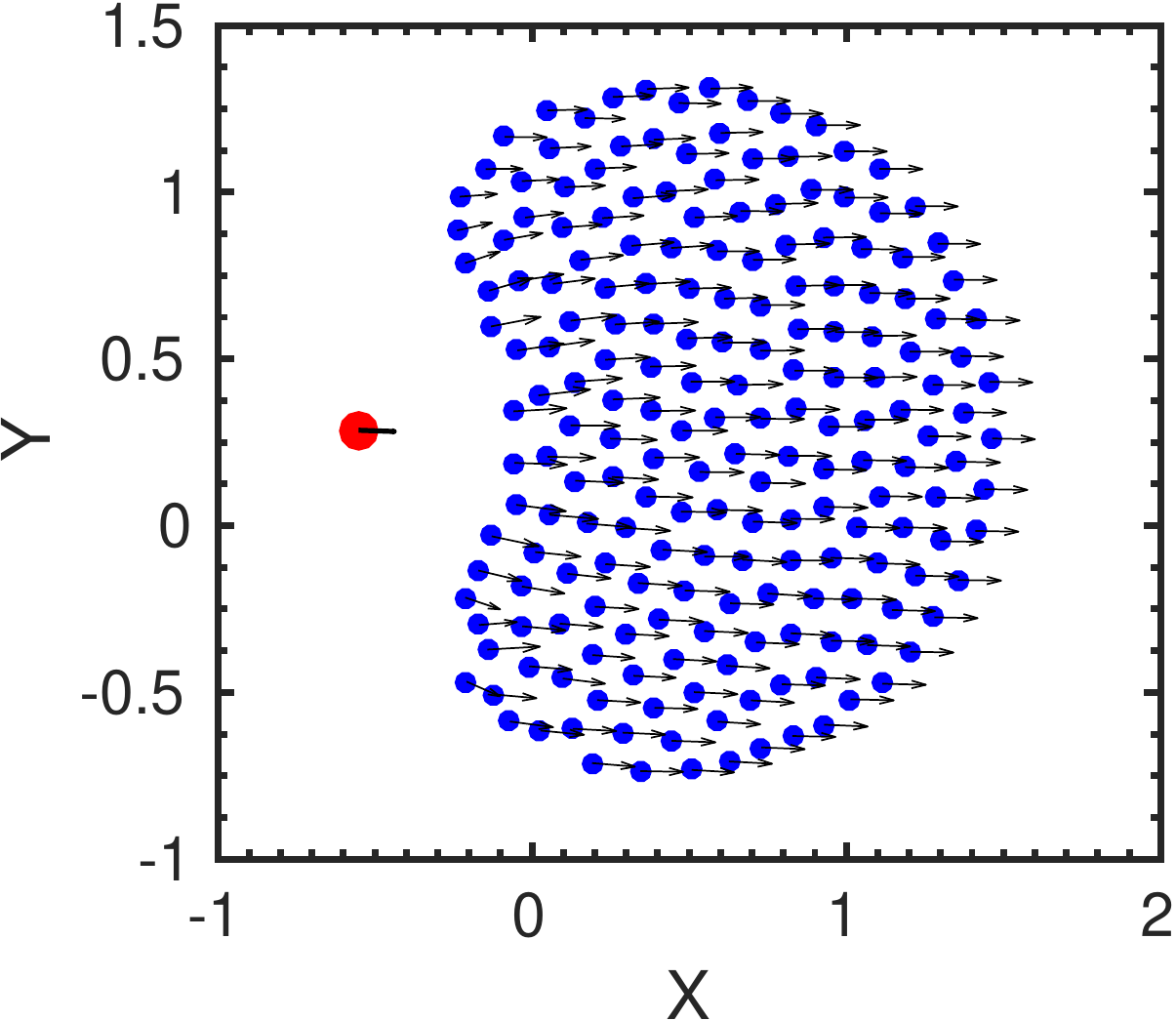}}
  	\subfigure[ ]{\label{fig:mpr_1pt0_mpd_3pt0_t320}\includegraphics[width=3.5cm,height=3.5cm]{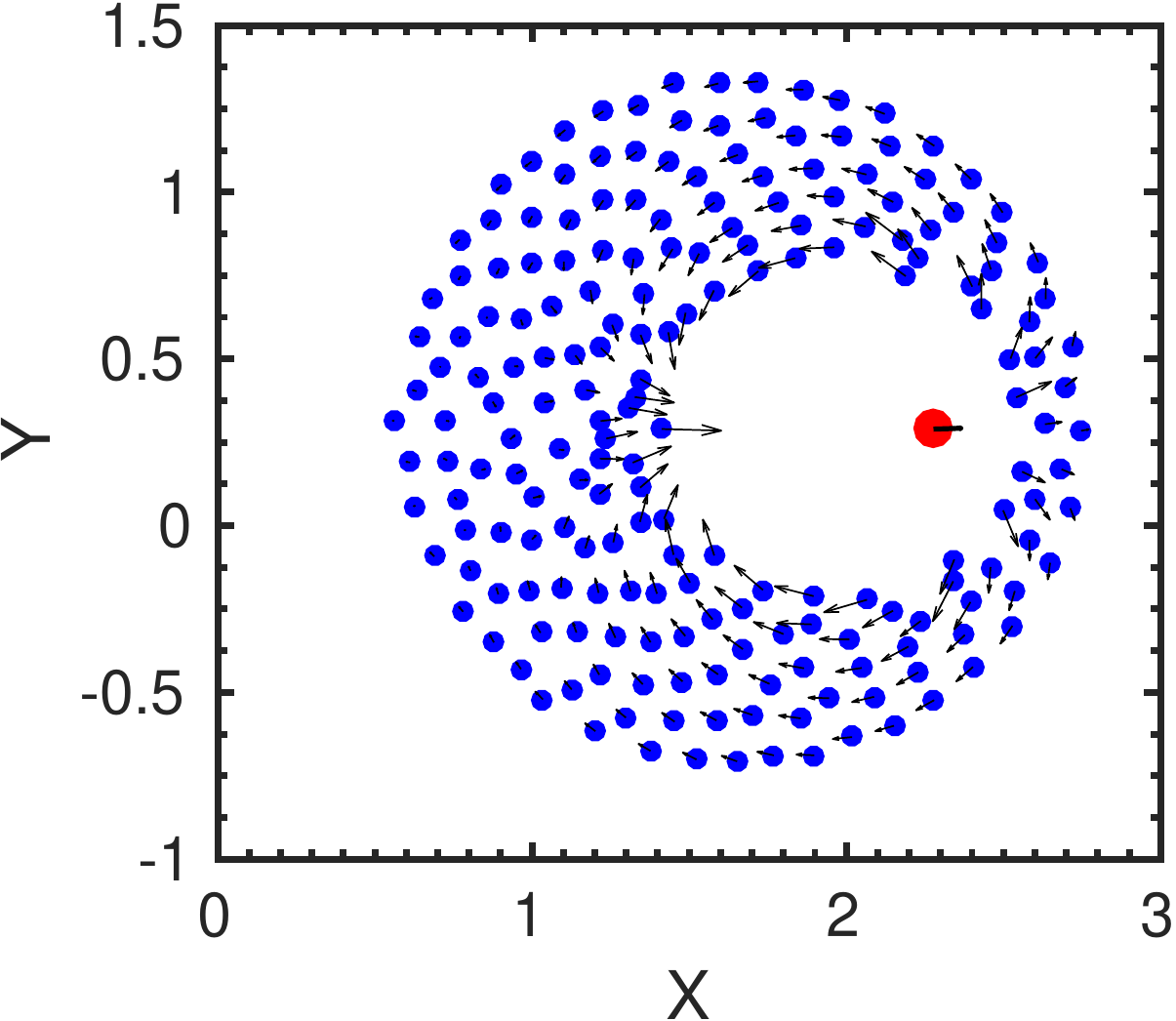}}
  	\subfigure[ ]{\label{fig:mpr_1pt0_mpd_3pt0_t330}\includegraphics[width=3.5cm,height=3.5cm]{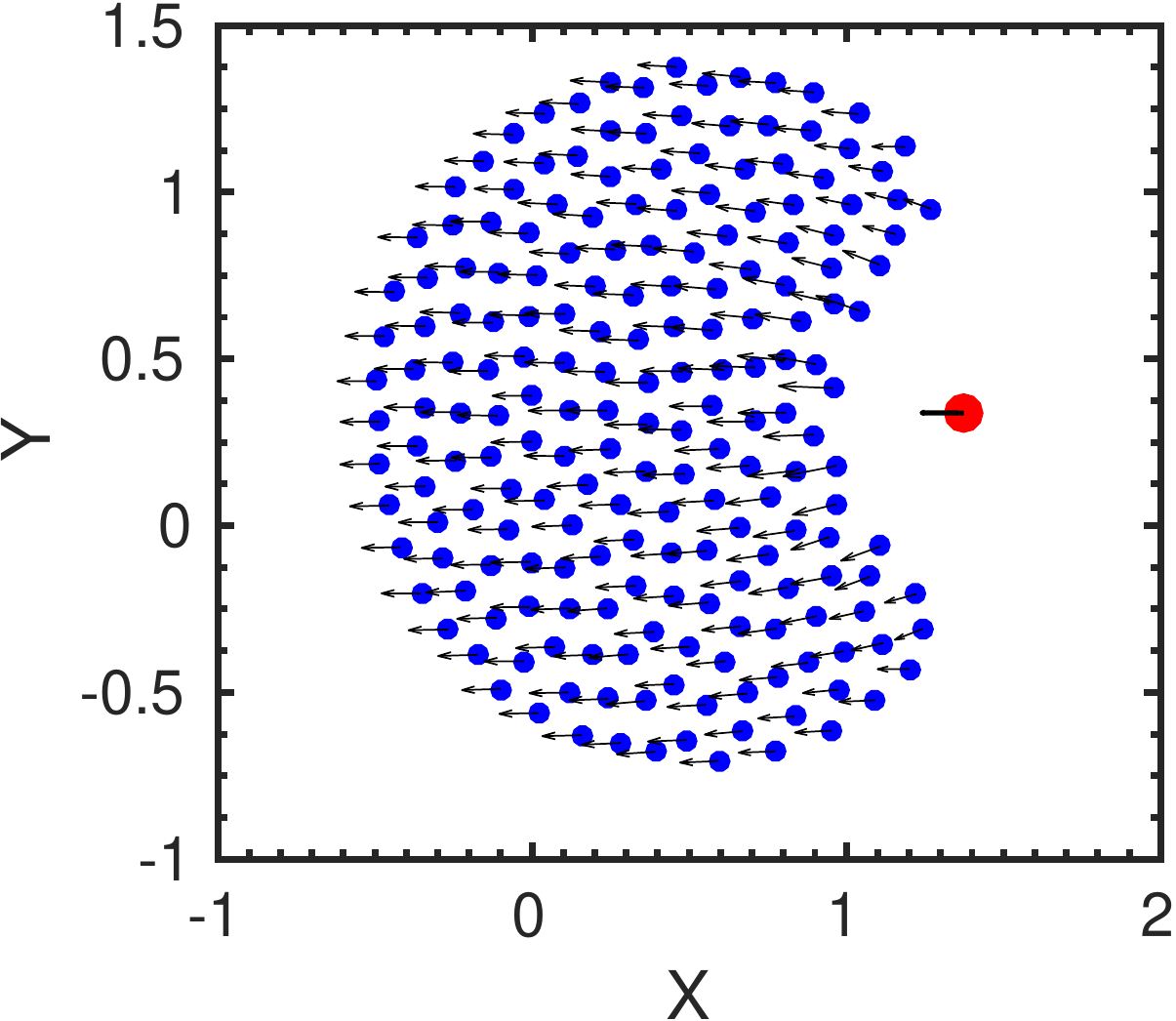}}
  	
  	\caption{Escape patterns of a prey swarm under the predator attack (predator strength, $\delta_0=0.4$). (a) Shows a stable ring formation by the prey swarm around the predator where the predator stays confused at the center, at simulation time $T=400$ with the predator mass, $M_{\rm pd}=2.0$ and prey mass, $M_{\rm pr}=1.0$. (b, c, d, e) Demonstrate the transition from stable ring state to chasing dynamics at time $T=300$, $310$, $320$, and $330$ with $M_{\rm pd}=3.0$ and $M_{\rm pr}=1.0$, (prey are shown by blue dots and the predator by red dot). }
  	
  	\label{figure_2}
\end{figure*}
We have solved the coupled Eqs. (\ref{preyequation}) and (\ref{predequation}) for a group of $N$ prey chased by a nearby predator. Here, we have explored how the inertial forces affect the survival and hunting dynamics of the prey swarm and the predator. The results are presented with varying the prey and predator mass, $M_{\rm pr}$ and $M_{\rm pd}$, and for different predator strength, $\delta_0$.
The other parameters are kept constant at $\alpha_0=1.0$, $\beta_0=1.0$, $\gamma_0=0.2$, and the kill radius $=0.01$.

\begin{figure}[!b]
	\centering  		
	  	\subfigure[ ]{\label{schematic}\includegraphics[width=3.5cm,height=3.5cm]{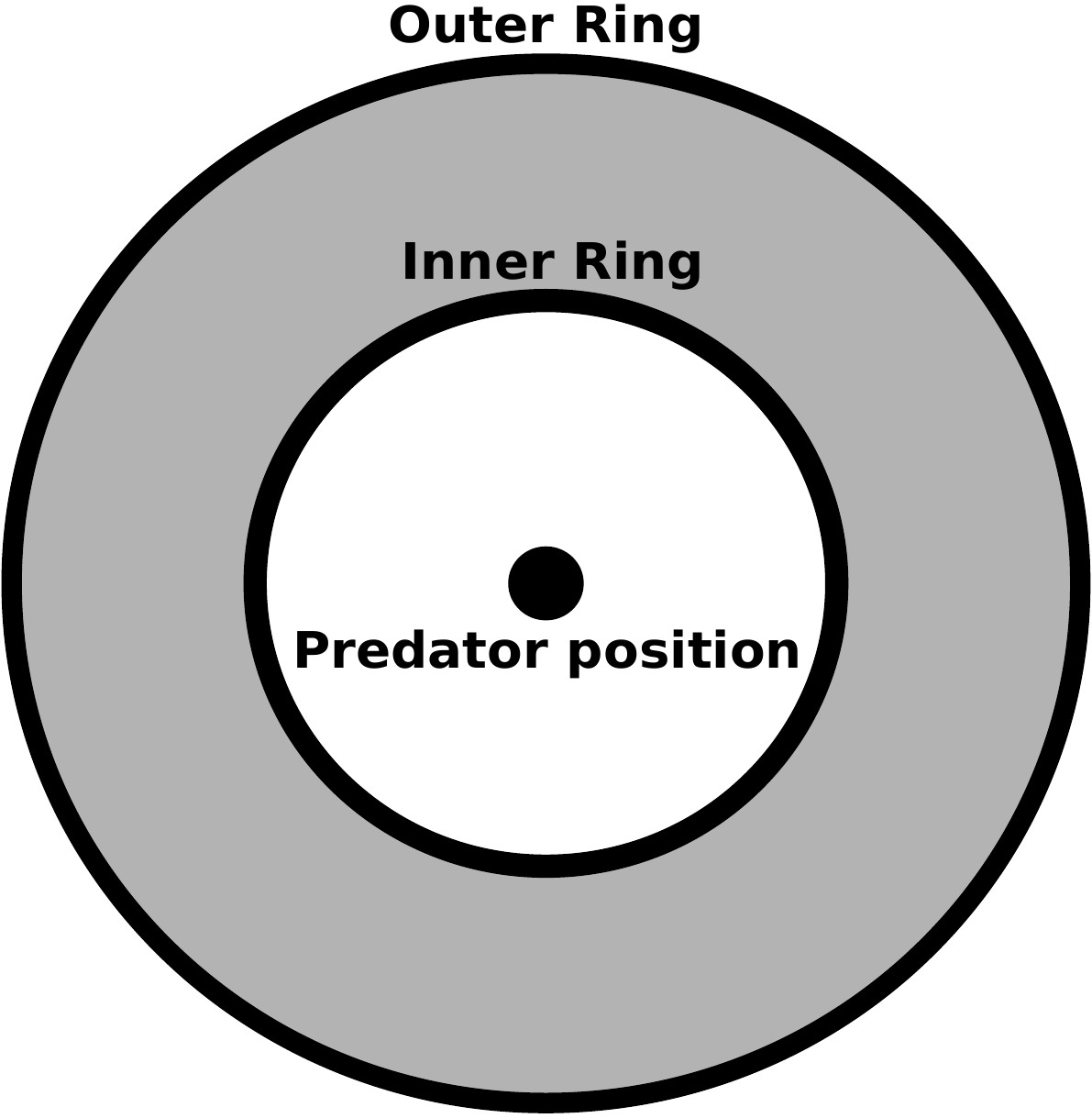}}
	  	\hspace{0.2cm}
	  	\subfigure[ ]{\label{eigenvalues}\includegraphics[width=4.2 cm,height=4.1cm]{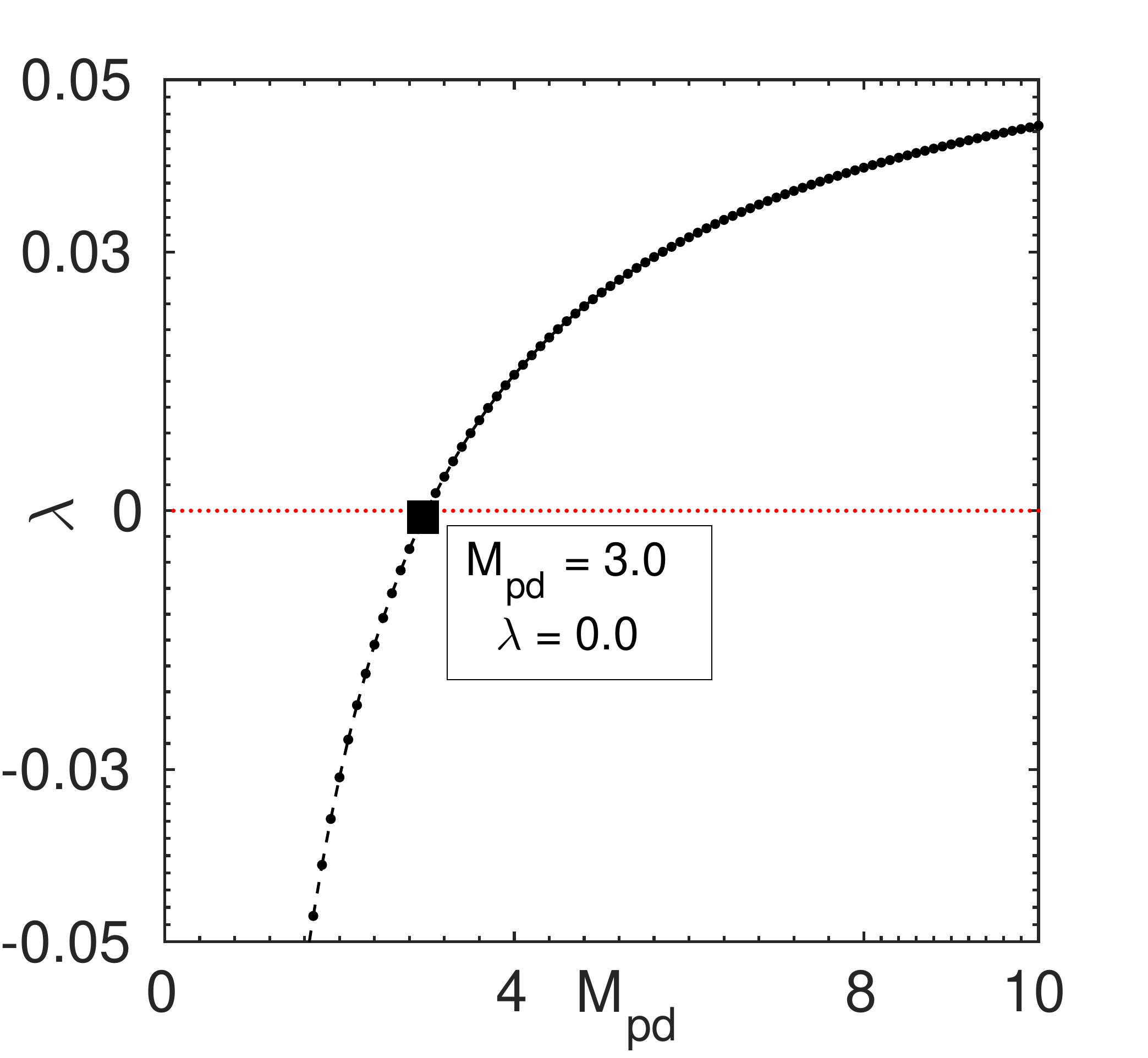}}
		\caption{(a) Presents a schematic of the steady state configuration where the prey swarm forms a stable circular ring around the predator. (b) Shows the evolution of one of the eigenvalues, $\lambda$, of the stability matrix (Eq. \ref{matrix_equation} in Appendix) with increasing predator mass, $M_{\rm pd}$, which crosses zero value at $M_{\rm pd} = 3.0$ (keeping $M_{\rm pr} = 1.0$ and other parameter values as in Fig. \ref{figure_2}).}
	\label{figure_3}
\end{figure}

We first investigate how the mass of the prey and predator influence the dynamics in the case of a weak predator when the whole prey group can survive the predator attack.
Figure \ref{figure_2} shows some representative simulation results with $\delta_0=0.4$, indicating a weak predator case.
We find that at steady-state, the prey
swarm forms a circular ring around the predator for a lower regime of predator mass.
Figure  \ref{fig:mpr_1pt0_mpd_2pt0_t400} presents such stable ring formation at $M_{\rm pd}=2.0$ keeping the value of $M_{\rm pr}=1.0$.
As we increase the mass of the predator to a higher value ($M_{\rm pd}> 3.0$), keeping all other
parameters fixed, the ring gets unstable, and the chasing dynamics emerge, as seen from Figs. \ref{figure_2}(b)-(e).  
These findings are also substantiated by analytical calculations and 
the inner radius of the ring,
$R_1=\sqrt{\frac{\gamma_0}{\beta_0}}$, and the outer radius of the ring, $R_2=\sqrt{\frac{1 + \gamma_0}{\beta_0}}$ match well with the numerical results presented in Fig. \ref{figure_2}(a)(the details are given in Appendix).
Further, we have performed a linear stability analysis of
the steady-state configuration by considering the perturbations at the inner, outer, and the predator
boundary as depicted in Fig. \ref{figure_3}(a). 
A detailed calculation shows that one of the eigenvalues of the stability matrix goes through a transition from a negative to positive value as the predator mass increases (keeping the prey mass fixed). 
The eigenvalue plotted in Fig. \ref{figure_3}(b) shows such transition at $M_{\rm pd}=3.0$
(plugging the numerical values as in Fig. \ref{figure_2}) which suggests that the system goes from a stable  state to an unstable
state with increasing predator mass, as observed in simulations, Fig. \ref{figure_2}.
This shows that inertia has a profound effect in
determining the steady-state configurations.
 
\begin{figure*}[!t]
 	\subfigure[ ]{\label{fig:a}\includegraphics[width=3.5cm,height=3.5cm]{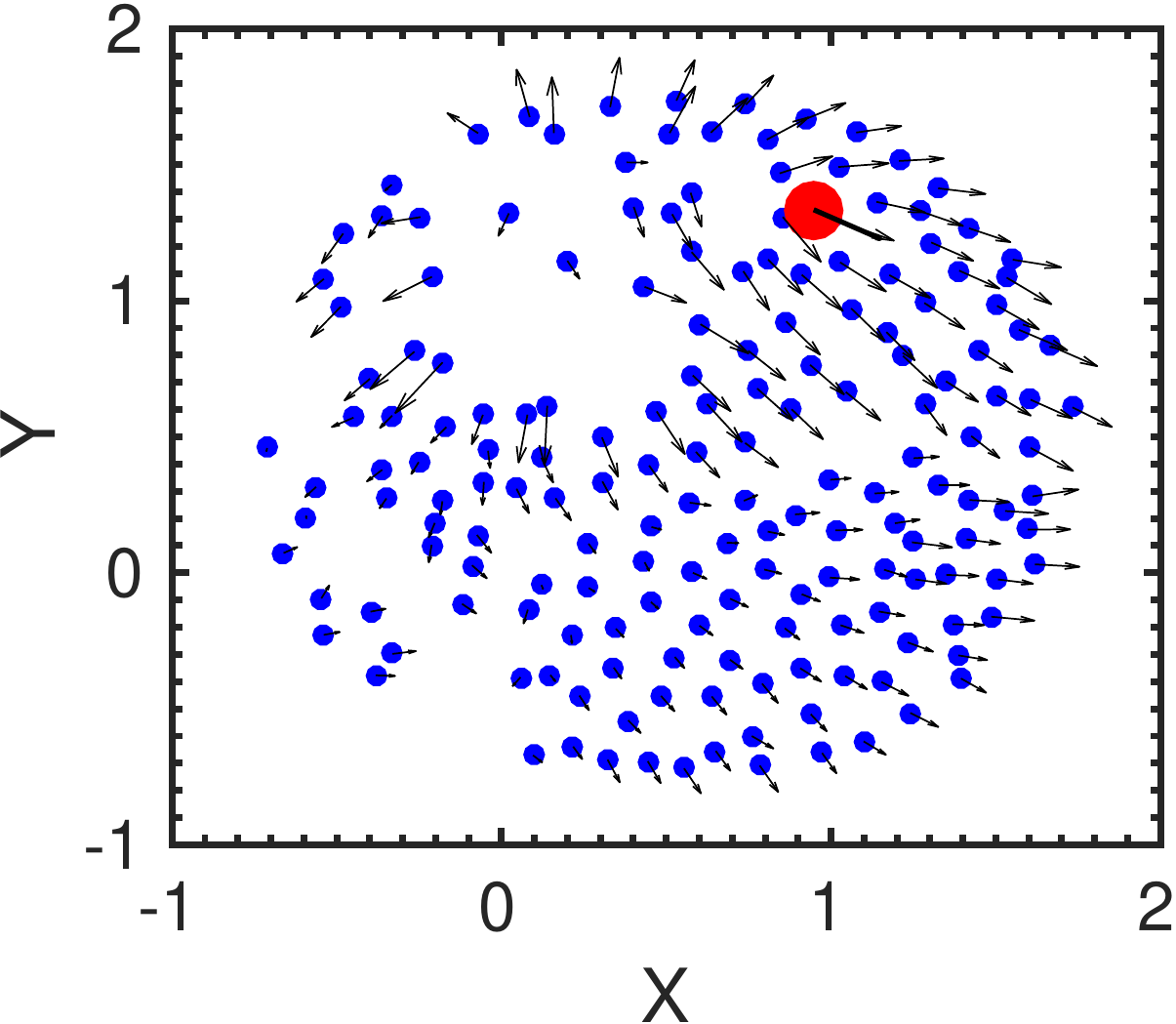}}
 	\hspace*{0.1cm}
 	\subfigure[ ]{\label{fig:b}\includegraphics[width=3.5cm,height=3.5cm]{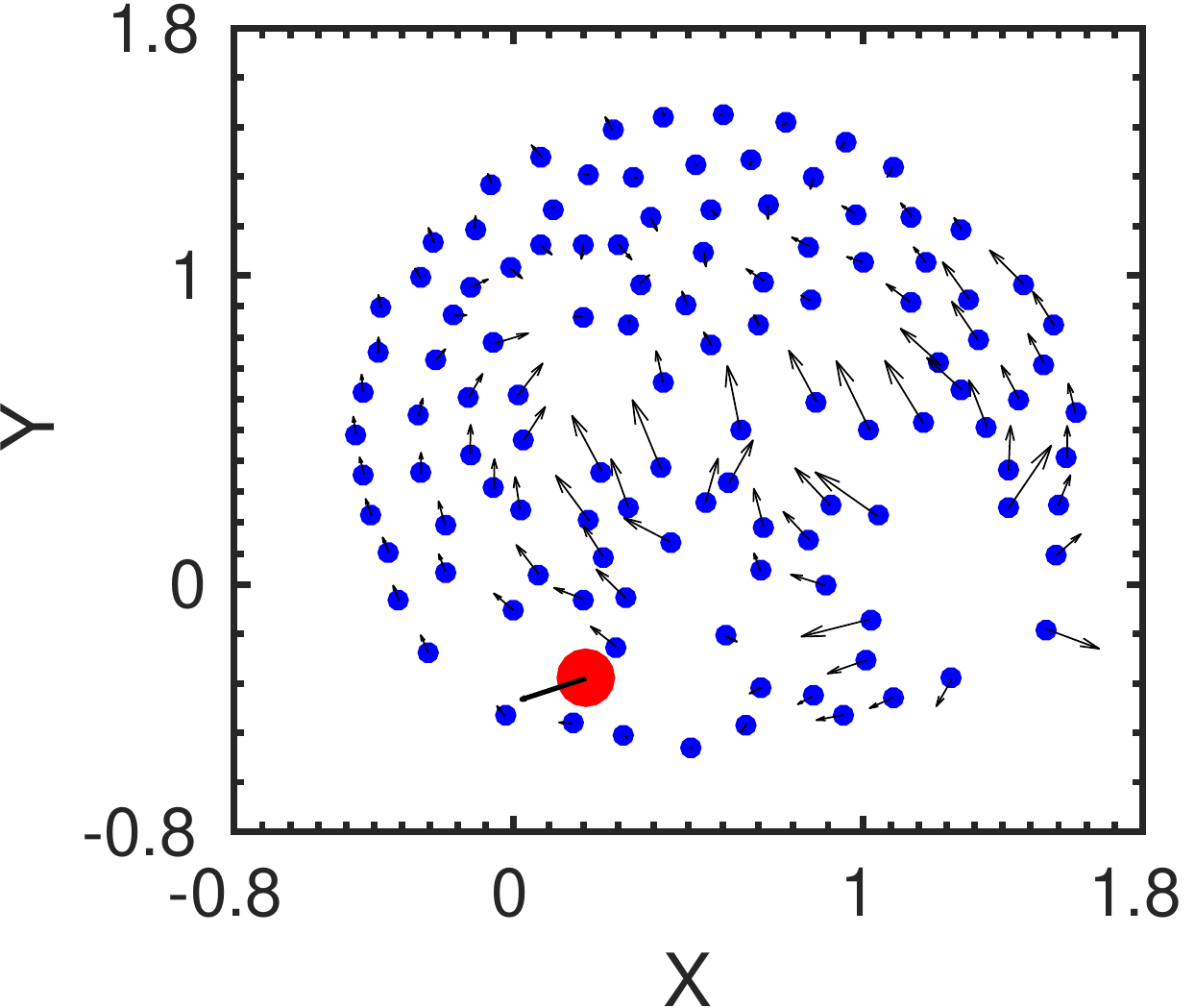}}
 	\hspace*{0.1cm}
 	\subfigure[ ]{\label{fig:c}\includegraphics[width=3.5cm,height=3.5cm]{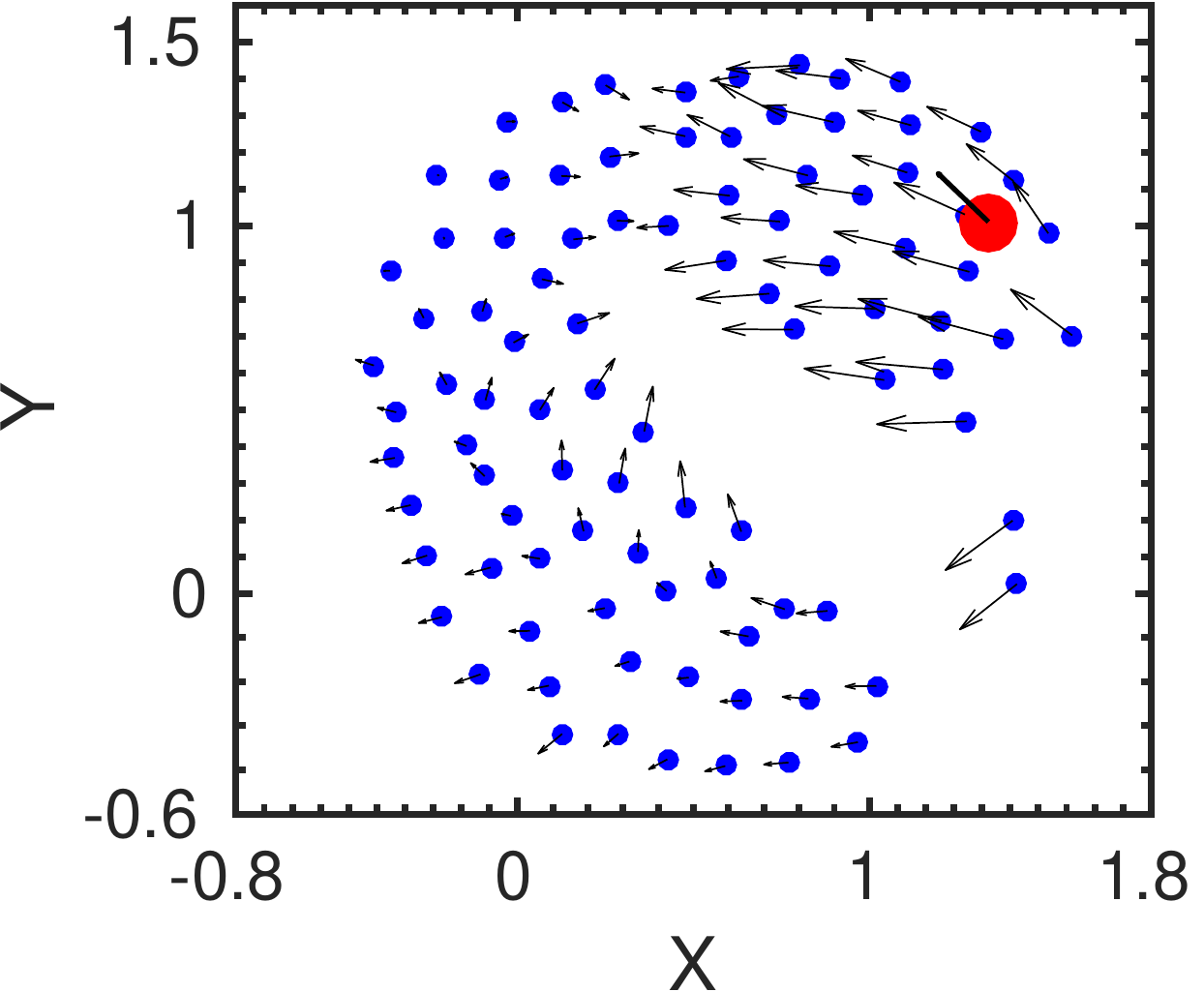}}
 	\hspace*{0.1cm}
 	\subfigure[ ]{\label{fig:d}\includegraphics[width=3.5cm,height=3.5cm]{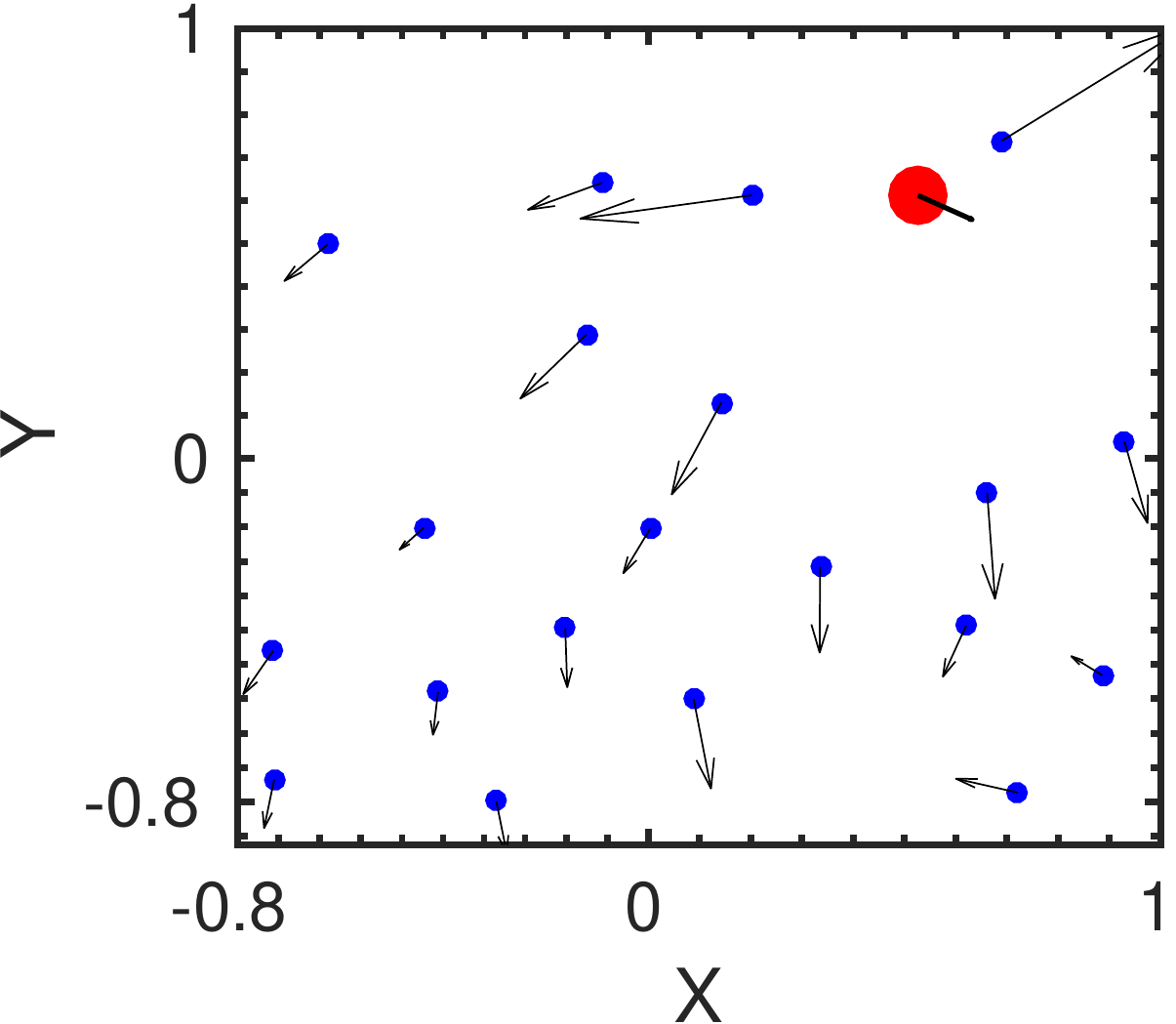}}
 	
 	\subfigure[ ]{\label{fig:e}\includegraphics[width=3.5cm,height=3.5cm]{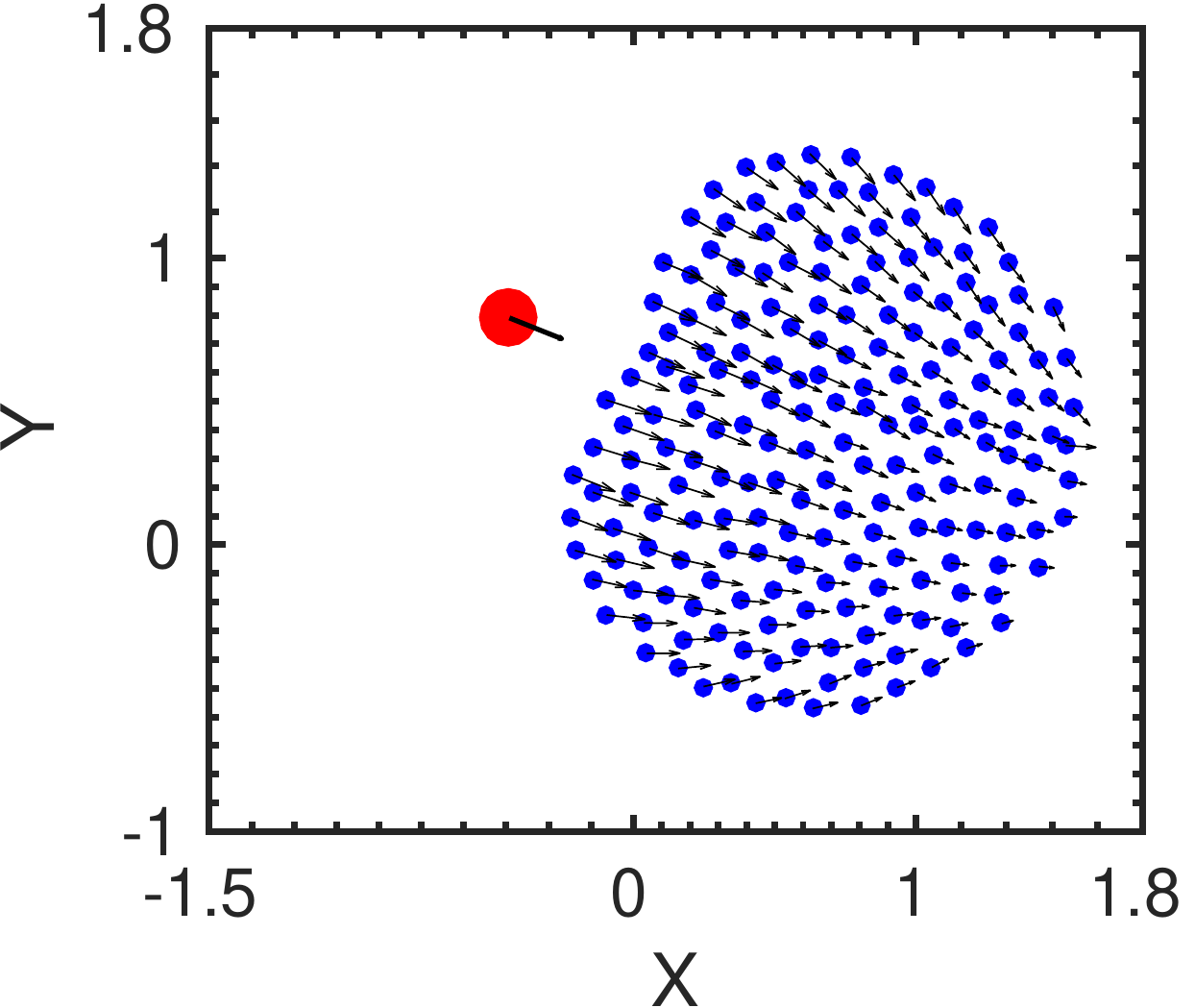}}
 	\hspace*{0.1cm}
 	\subfigure[ ]{\label{fig:f}\includegraphics[width=3.5cm,height=3.5cm]{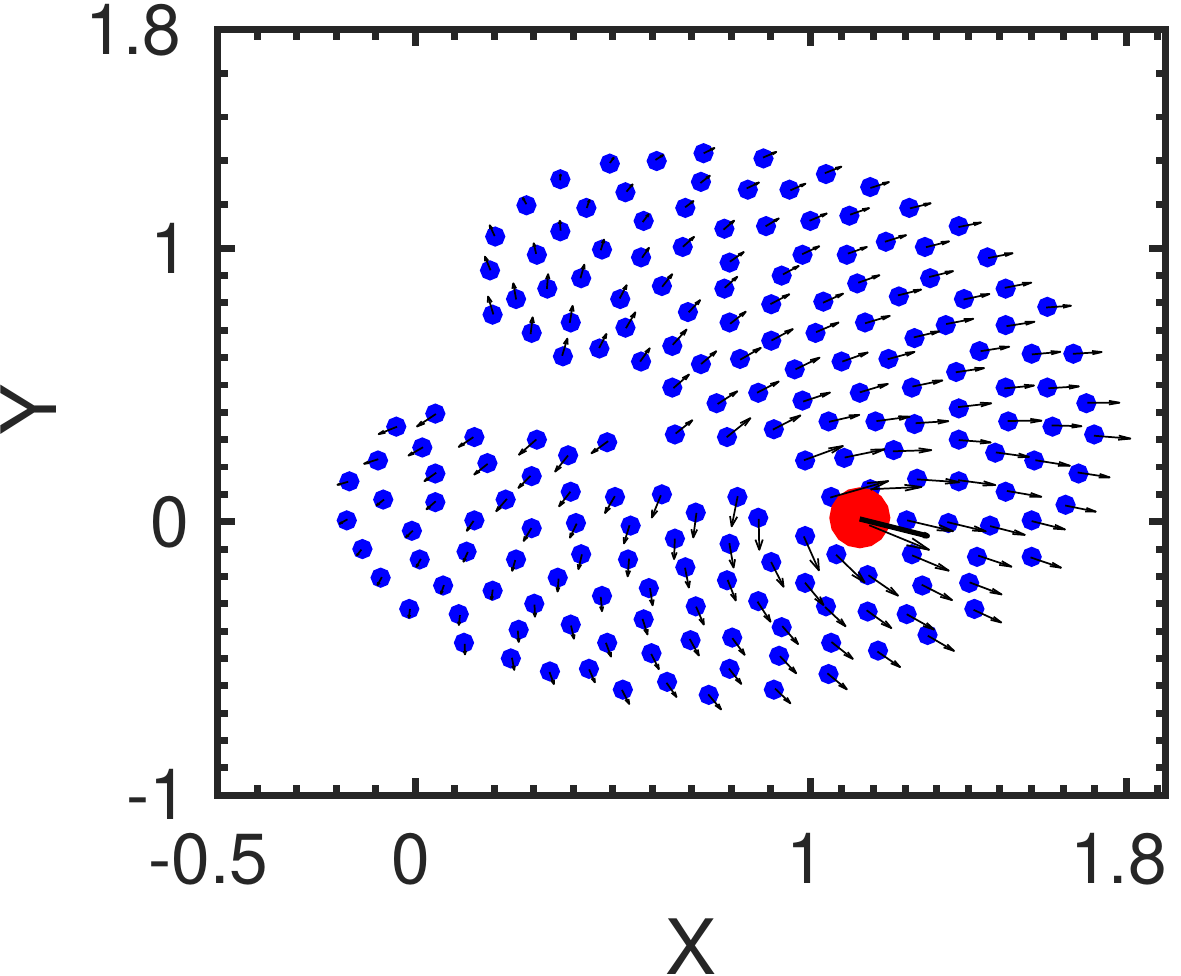}}
 	\hspace*{0.1cm}
 	\subfigure[ ]{\label{fig:g}\includegraphics[width=3.5cm,height=3.5cm]{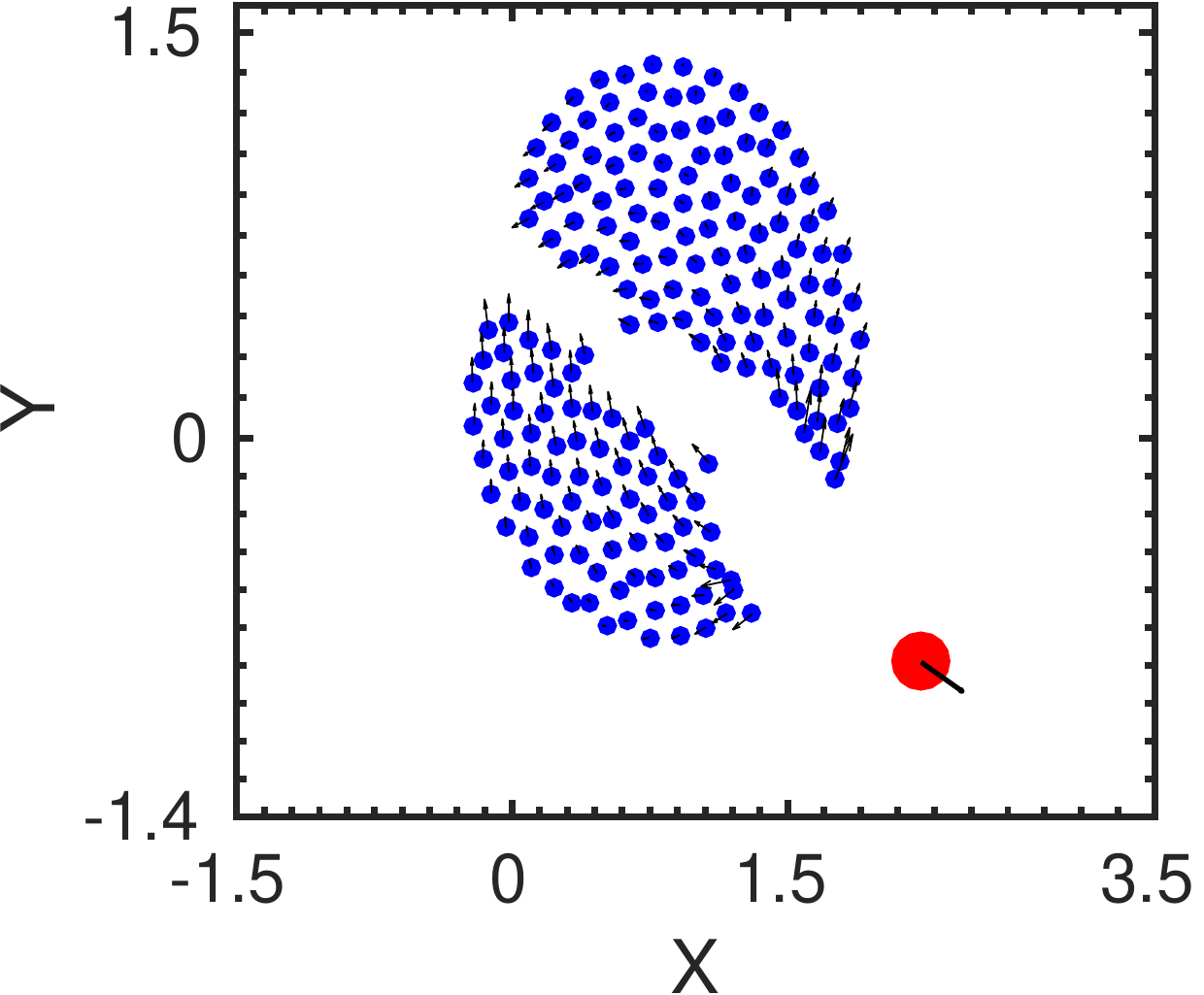}}
 	\hspace*{0.1cm}
 	\subfigure[ ]{\label{fig:h}\includegraphics[width=3.5cm,height=3.5cm]{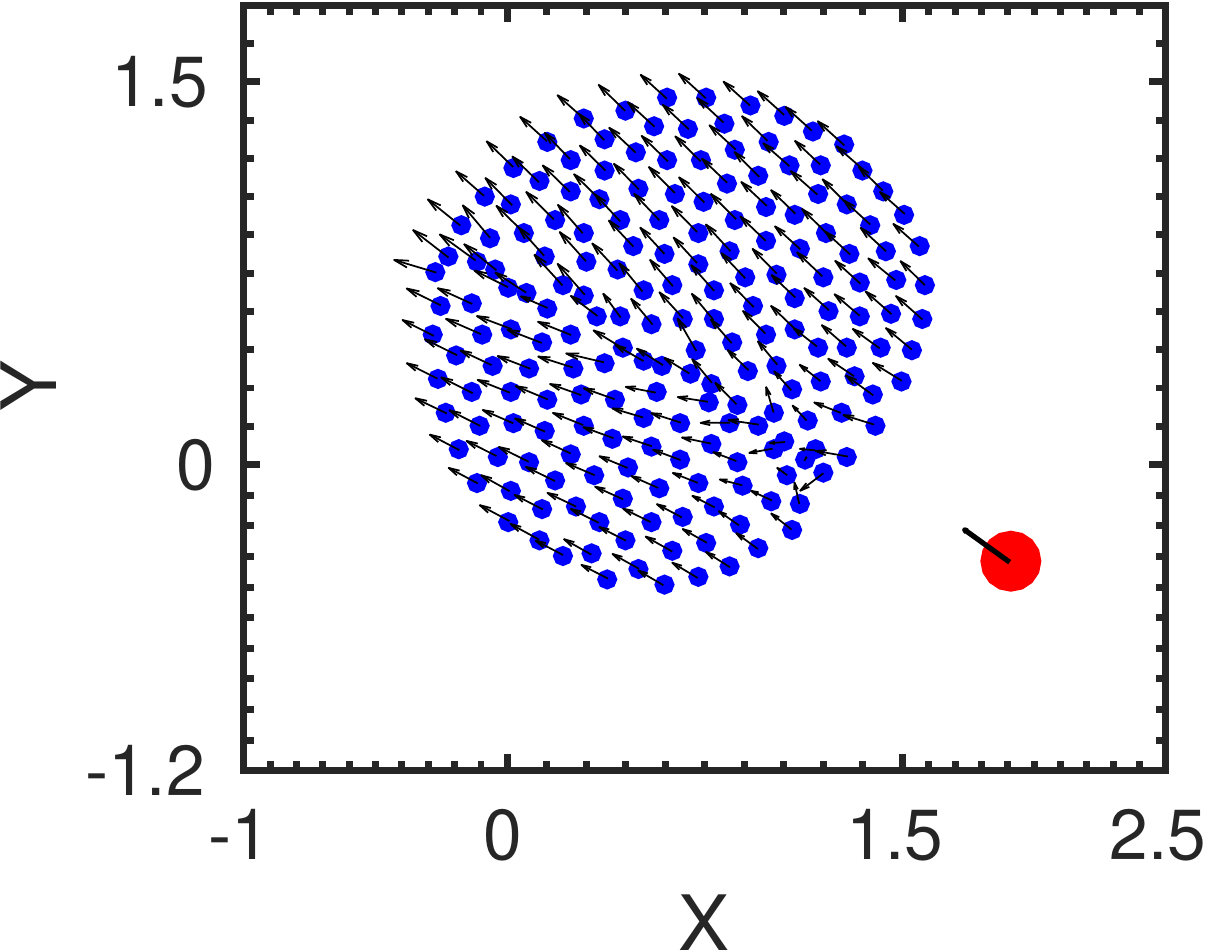}}
 	
 	\subfigure[ ]{\label{fig:i}\includegraphics[width=3.5cm,height=3.5cm]{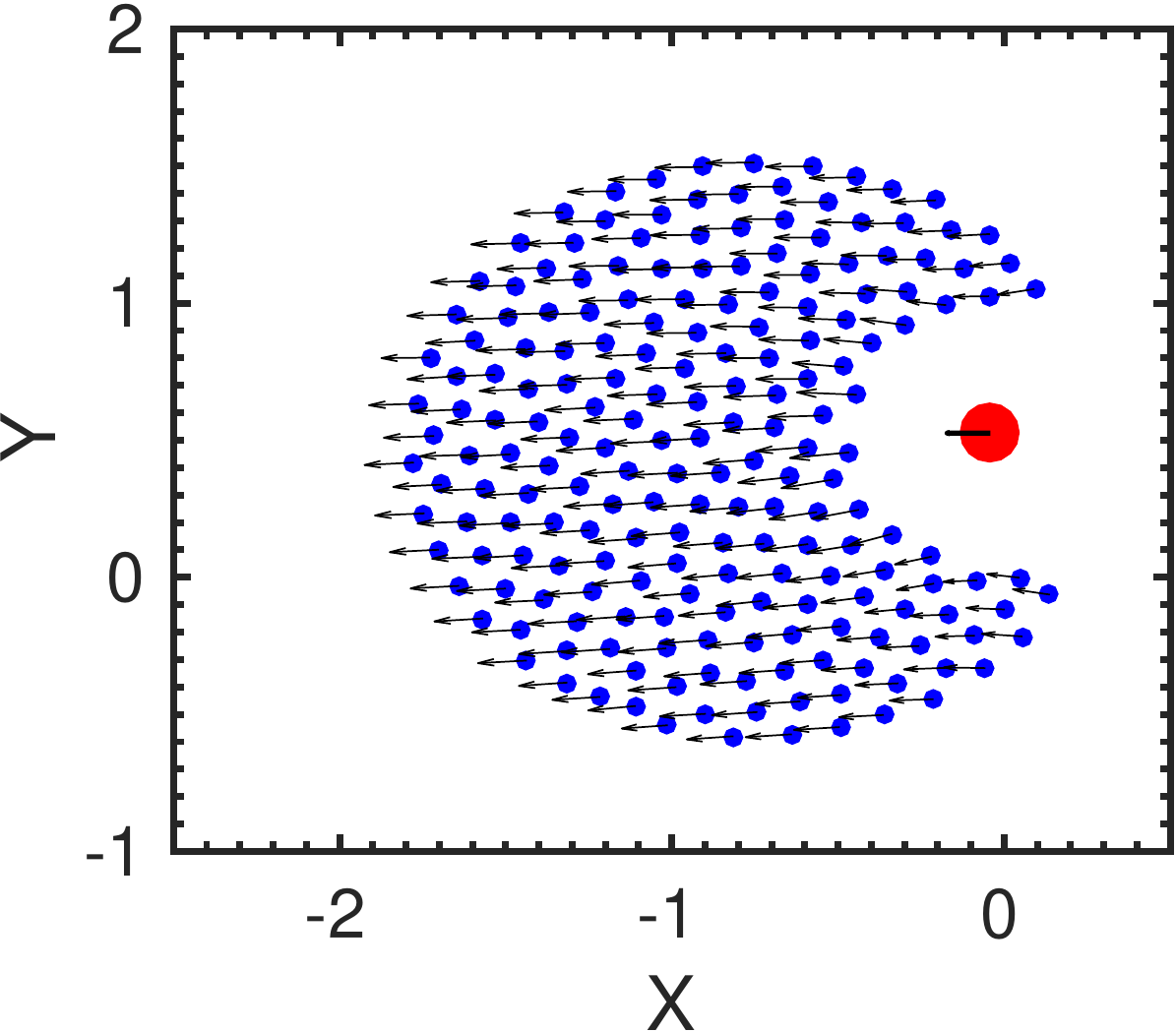}}
 	\hspace*{0.1cm}
	\subfigure[ ]{\label{fig:j}\includegraphics[width=3.5cm,height=3.5cm]{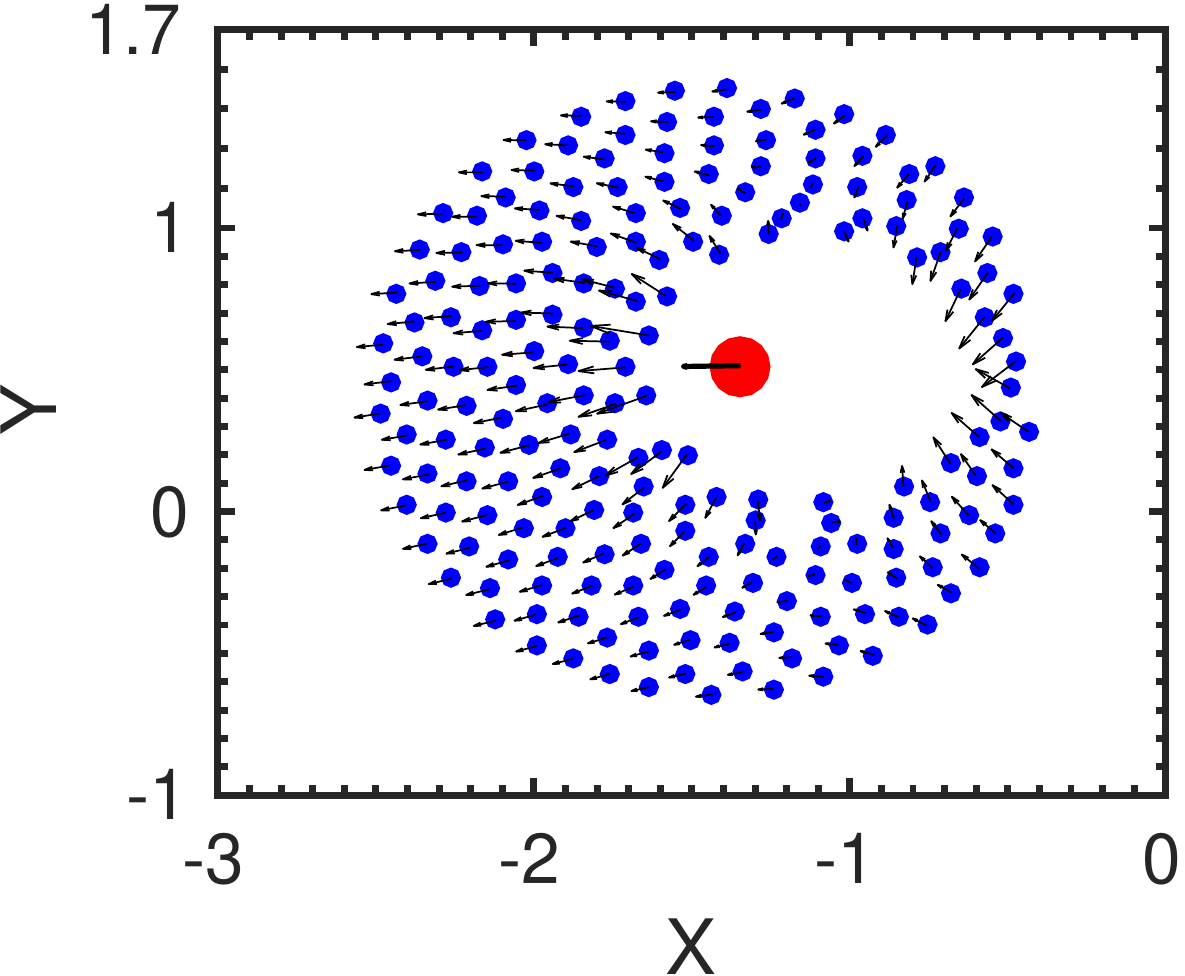}}
	\hspace*{0.1cm}
	\subfigure[ ]{\label{fig:k}\includegraphics[width=3.5cm,height=3.5cm]{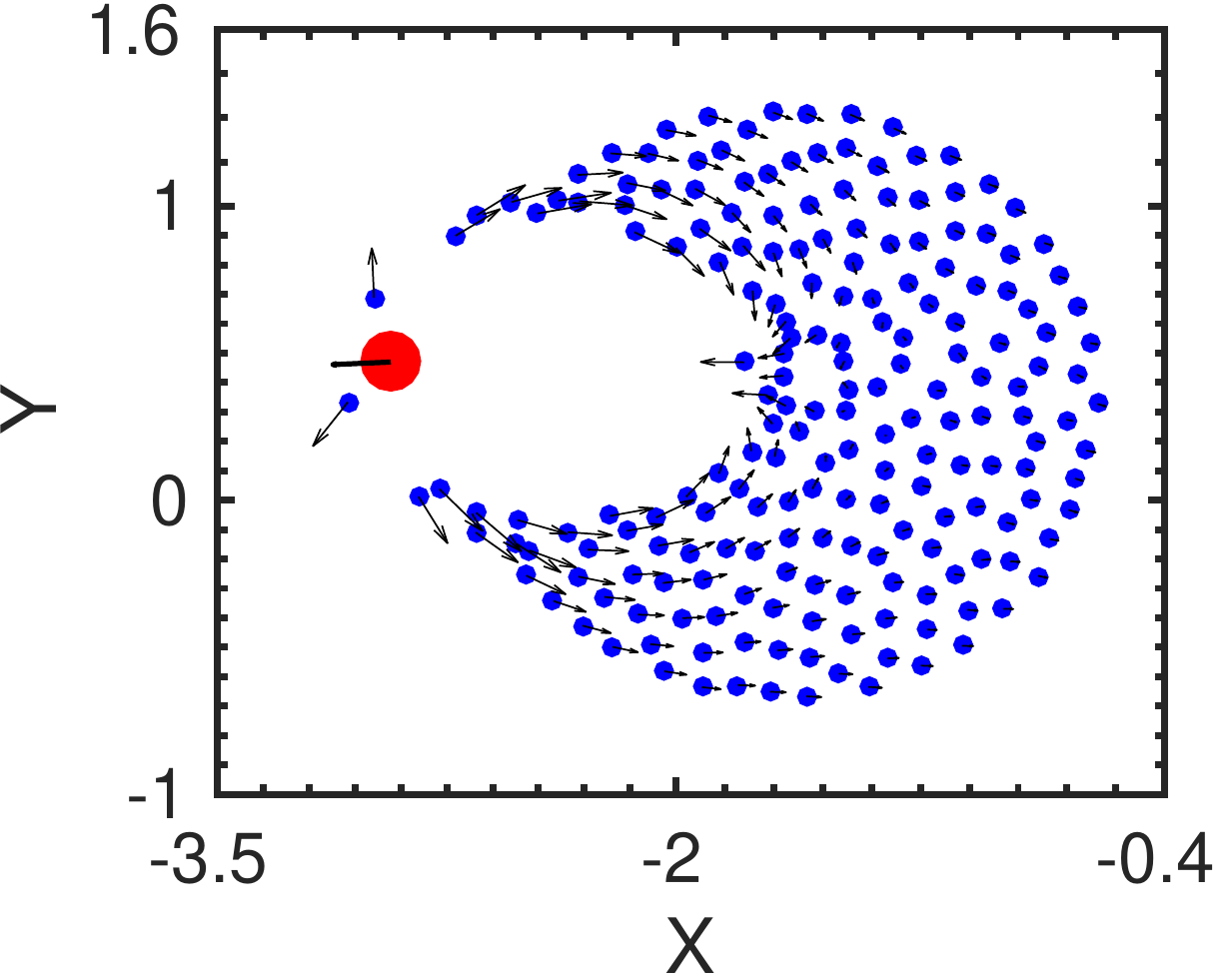}}
	\hspace*{0.1cm}
	\subfigure[ ]{\label{fig:l}\includegraphics[width=3.5cm,height=3.5cm]{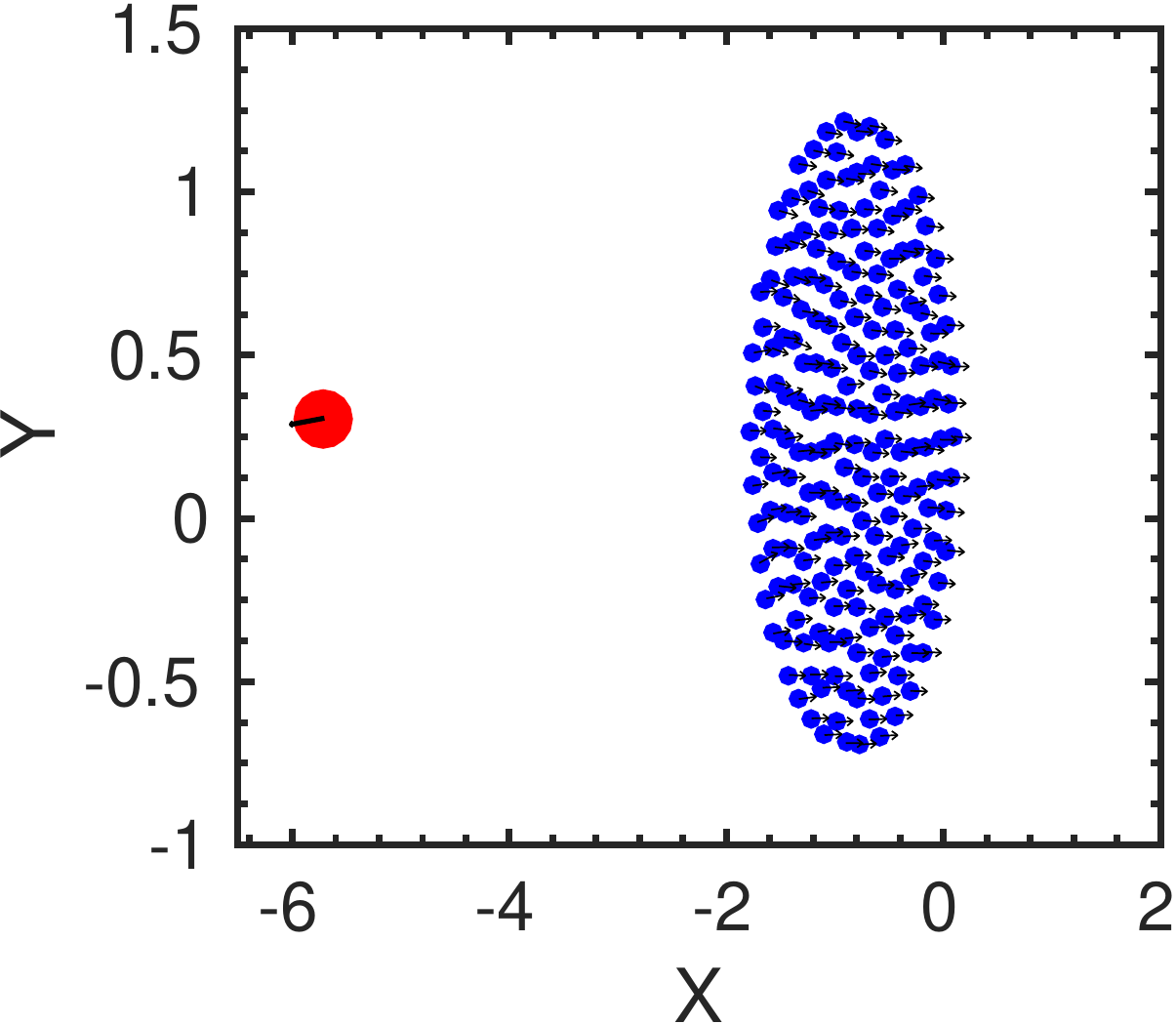}}
 	
 	\caption{ 
 		Various emergent escape trajectories have been shown for different values of $M_{\rm pd}$ keeping $M_{\rm pr}=1.0$. (a, b, c, d) Show the snapshots of chasing dynamics and capture by the predator at different simulation time, $T=2, 10, 15, 30$ for $M_{\rm pd}=0.1$, {\it i.e.}, for lighter predator mass. (e, f, g, h) Prey swarm divides into subgroups and merges again into a unitary group to avoid the predator attack for slightly heavier predator mass $M_{\rm pd}=1.0$; snapshots are at $T=4.0, 5.0, 6.0, 9.0$. (i, j, k, l) Exhibit the escape patterns for heavier predator mass ($M_{\rm pd}=100$). Here, the prey group turns back along an arc to confuse the predator and can easily escape away. The pattern is similar to the commonly observed F-maneuvering by the school of fishes; snapshots are at $T=10, 15.0, 20.0, 30.0$.} 
 	\label{phase_plot}
 	
 \end{figure*}

Next, we explore the effect of inertia in the strong predator regime, {\it i.e.}, when the value of $\delta_0$ is large.
We find the emergence of diverse escape patterns as the mass of the predator ($M_{\rm pd}$) and prey ($M_{\rm pr}$) are varied.
At a lower value of predator mass, $M_{\rm pd}=0.1$, keeping the prey mass at $M_{\rm pr}=1$ and $\delta_0=2.5$,  Figs. \ref{phase_plot}(a)-(d) show that the lighter predator continuously chases the prey group by turning its direction towards the prey when they try to escape away.
Eventually, the predator can capture the whole prey swarm after some time.
Now, as the predator mass is increased ($M_{\rm pd}=1.0$), {\it i.e.}, for a slightly heavier predator, a different escape trajectory emerges as shown in Figs. \ref{phase_plot}(e)-(h). As the predator approaches near to the prey group, the prey divides into two groups to avoid the predator; then, as the predator passes through the group, the divided prey groups again merge into a cohesive group, and this pattern repeats as time proceeds. This type of escape pattern is prevalent in the school of fishes defending the predator attack~\cite{brose2010funeco,pavlov2000JOI}. 
As the predator mass is increased to $M_{\rm pd}=100$,  the predator initially chases the prey swarm and moves in the same direction as the prey move, as in Fig. \ref{phase_plot}(i). After some time, the prey group divides and turns back along an arc to confuse the predator, as shown in Fig. \ref{phase_plot}(k). However, there is a delay in the predator's turning due to its heavier mass, and by that time, the prey group merges into a unitary group and escapes away from the predator as seen from Fig. \ref{phase_plot}(l).
This widespread predator avoidance has been observed in the school of fishes, and it is called \lq F-maneuver' or \lq Fountain effect' \cite{pavlov2000JOI}. 

Now, to characterize  the velocity patterns of the prey-predator, we define an order parameter  $\phi(T)$ to describe the average direction of motion of the prey group  with respect to the predator's direction as,
\begin{equation}
\phi(T)=\frac{1}{N_{\rm sur}}\sum_{i=1}^{N_{\rm sur}}{\rm cos(\theta_i)};
\end{equation} 
here $\theta_i$ is the angle between the velocity direction of  the i'th prey and the predator, which is calculated from,
 $\rm cos (\theta_i)=\frac{\vec{V}_{i,\rm prey}.\vec{V}_{\rm predator}}{|\vec{V}_{i,\rm prey}|.|\vec{V}_{\rm predator}|} $,  and $N_{\rm sur}$ is the number of survived prey at any time instant $T$. 
 
\begin{figure}[!t]
	\centering  		
	\subfigure[ ]{\includegraphics[width=4.0cm,height=4.0cm]{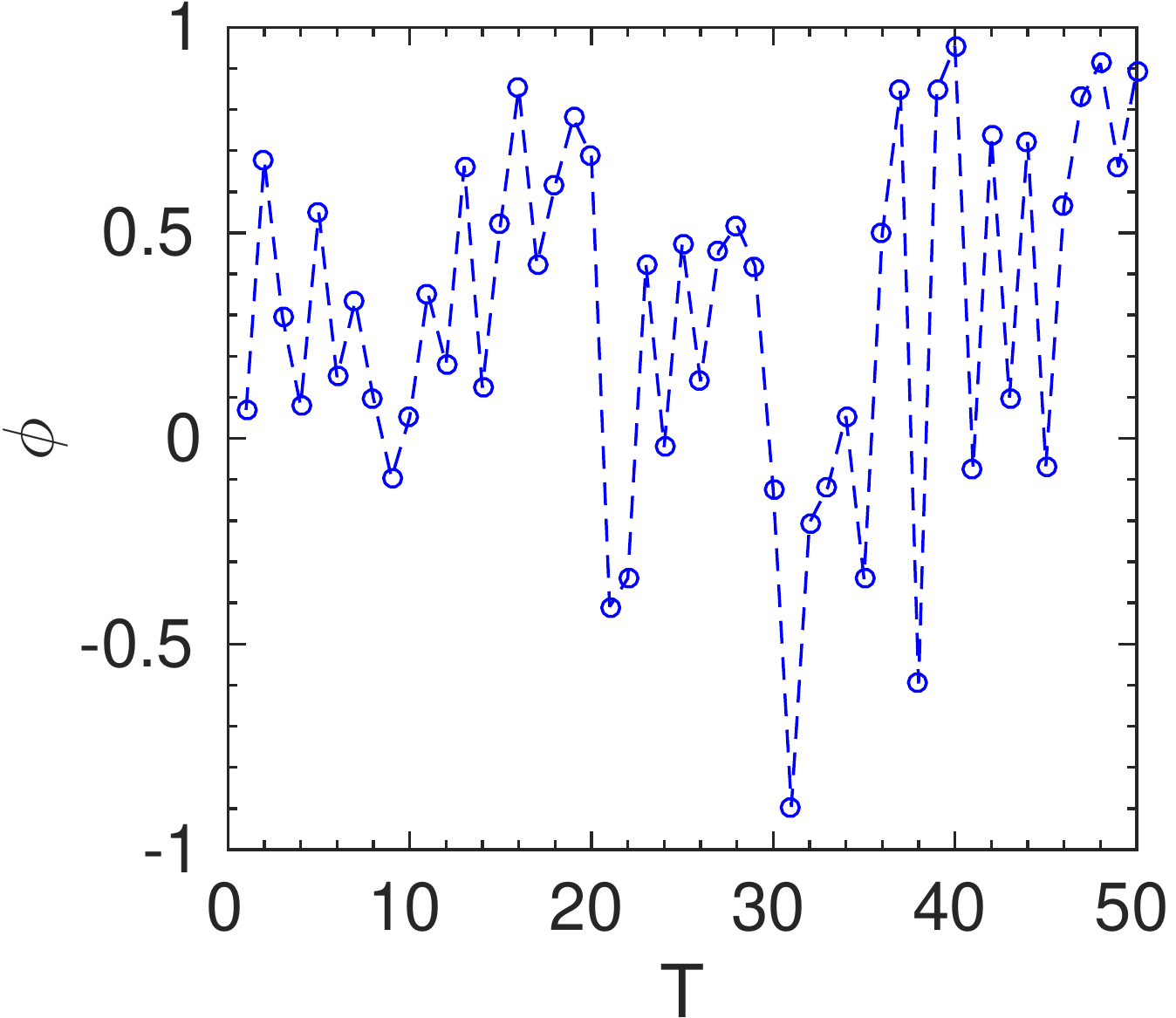}}
	\hspace{0.1cm}
	\subfigure[ ]{\includegraphics[width=4.0cm,height=4.0cm]{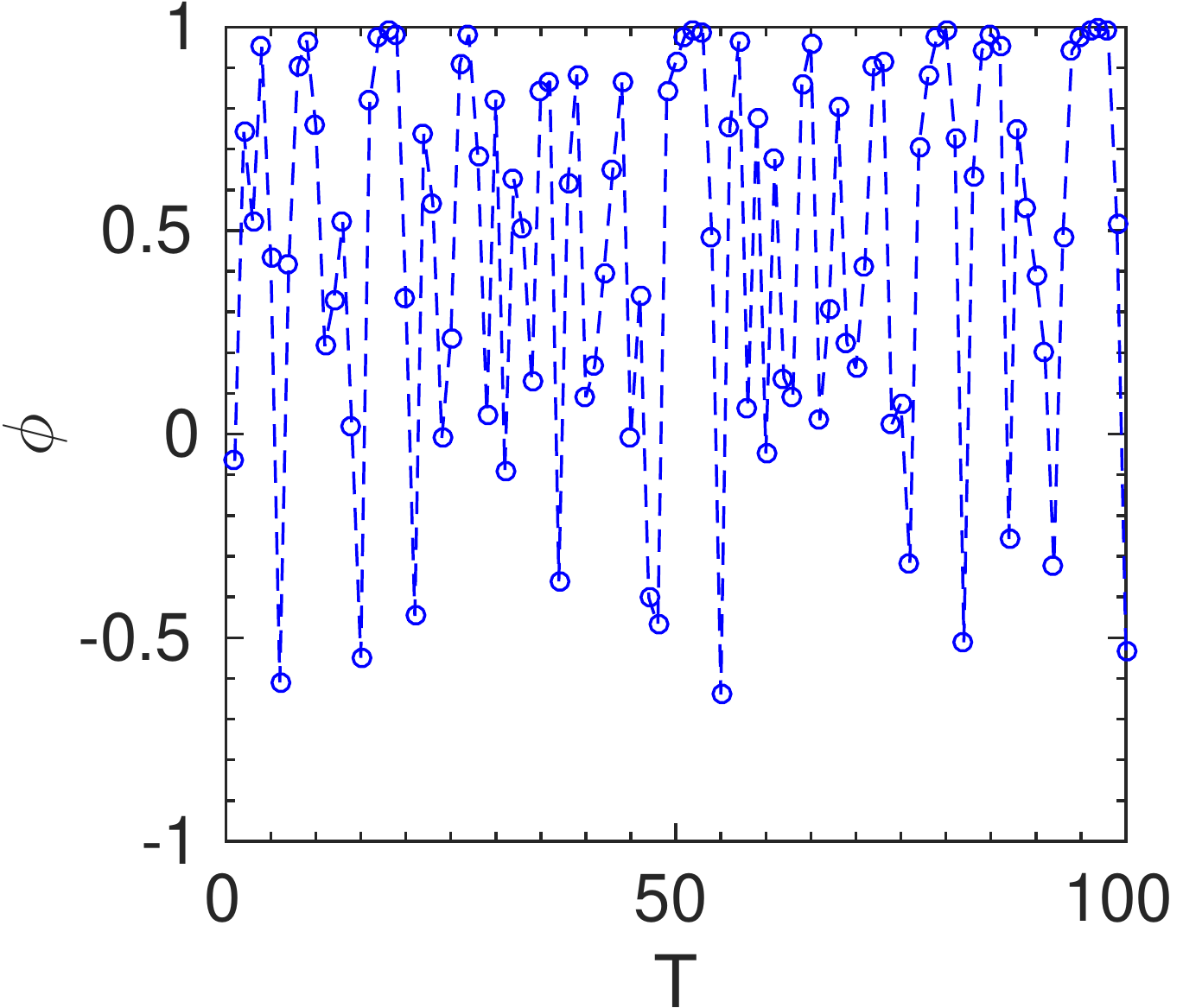}}
	\hspace{0.1cm}
	\subfigure[ ]{\includegraphics[width=4.0cm,height=4.0cm]{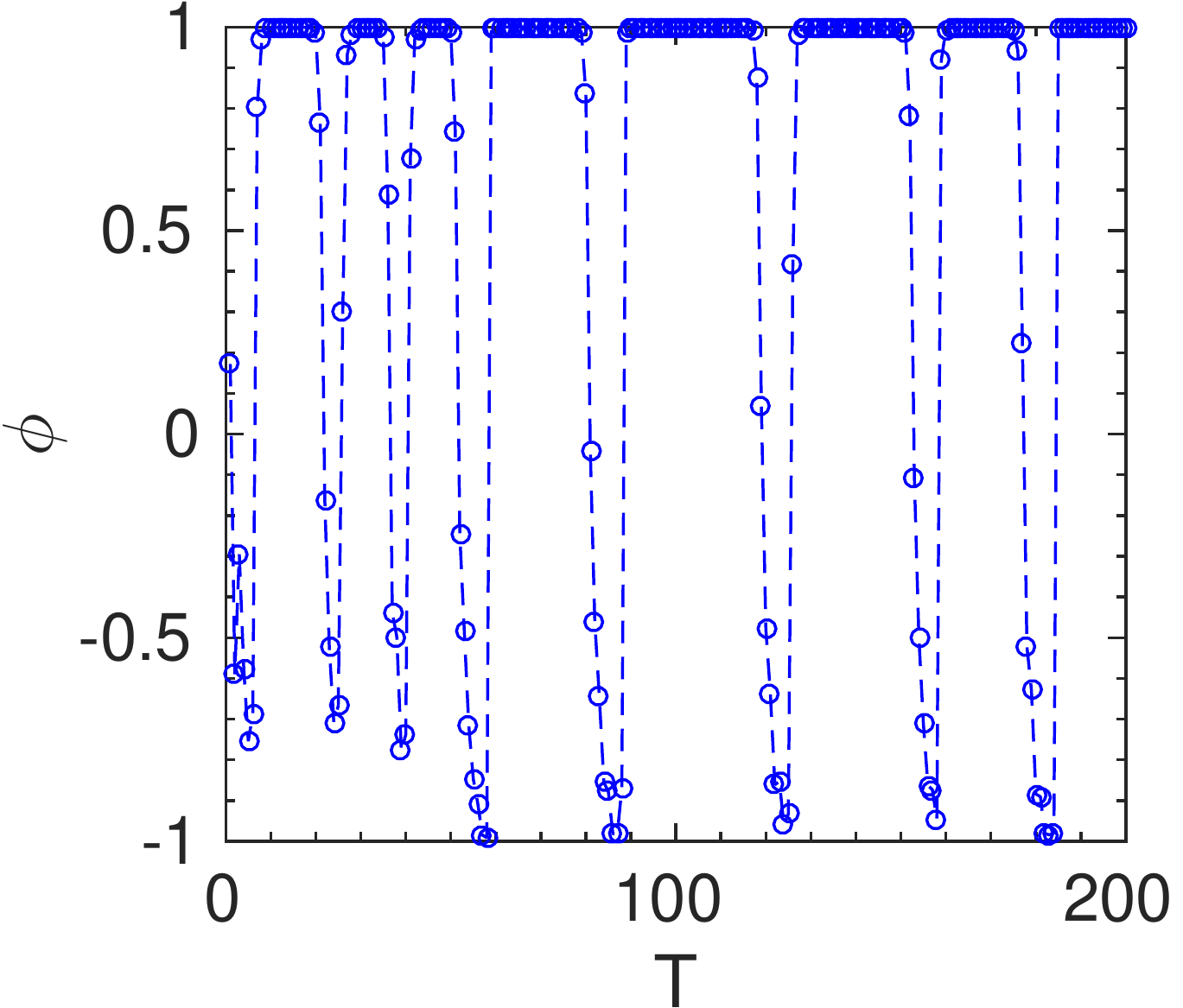}}
	\hspace{0.1cm}
	\subfigure[ ]{\includegraphics[width=4.0cm,height=4.0cm]{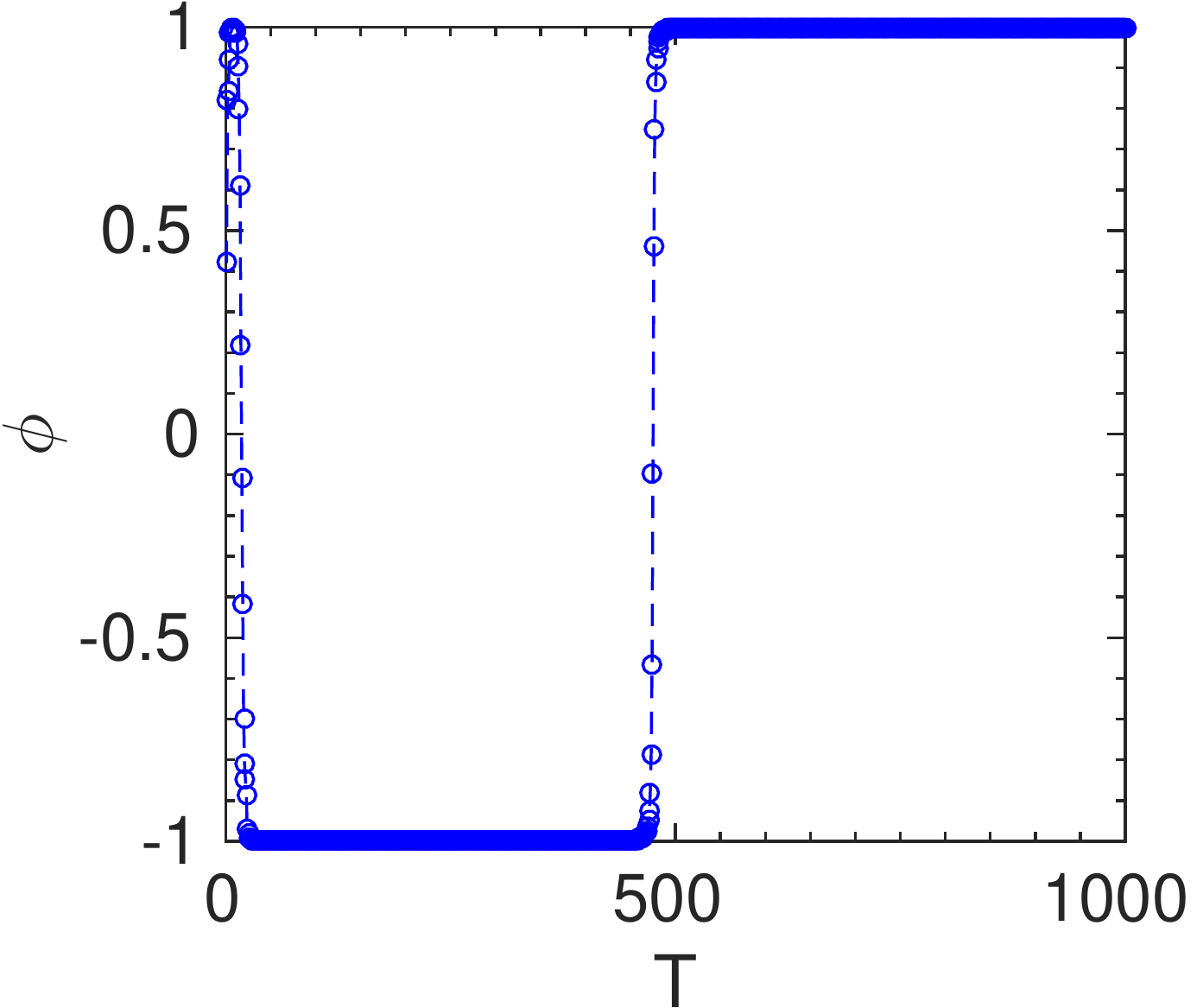}}
	\caption{Order parameter $\phi$ has been shown as a function of time, $T$, for different predator mass, $M_{\rm pd}=0.1,1.0,3.0$, and $100.0$ where $M_{\rm pr}=1.0$.}
	\label{order_param}
\end{figure}

 For a smaller predator mass, $M_{\rm pr}=0.1$, as seen from Fig. \ref{order_param}(a), $\phi$ changes rapidly because the predator continuously changes its directions to chase and capture the prey, and the prey also reorients themselves to survive the attack. 
With increase in the predator mass, $M_{\rm pr}=1.0$, as shown in Fig \ref{order_param}(b), $\phi$ value dynamically changes in between $1$ to $-1$.
When all prey collectively move in the same direction, and the predator approaches the group in that direction, as shown in the snapshot Fig. \ref{phase_plot}(e),
it gives rise to $\phi=1$.
However, the onset of prey-group division [as shown in Fig. \ref{phase_plot}(f)] leads to a lower value of $\phi$. Then, as the prey reorients and turns back in order to escape moving in the opposite direction of the predator [Fig. \ref{phase_plot}(g)], the value of $\phi$ approaches to $-1$. Eventually, the predator also reorients and chases the prey group in the same direction, Fig. \ref{phase_plot}(h). This pattern repeats as time progresses.
Hence, the $\phi$ value oscillates between  $1$ to $-1$. The time period of oscillation and the process of chase and capture is strongly dependent on the predator mass, $M_{\rm pd}$, as could be seen from Fig \ref{order_param}. 
With an increase in predator mass, $M_{\rm pd}=3.0$ shown in Fig. \ref{order_param}(c), the predator requires a longer time to reach closer to the prey swarm, and thus $\phi$ value stays at $1$ for a longer duration. In the case of the very high mass of the predator, $M_{\rm pd}=100.0$, the prey group could efficiently escape from the predator, shown in Fig. \ref{phase_plot}(i)-(l).
Due to the higher mass, the predator takes a much longer time to turn back towards the prey group to chase them again; hence $\phi$ stays at $-1$ for a very long time, as could be seen from Fig. \ref{order_param}(d).
 By the time the prey group goes far away, so even though the predator chases the group but it can not catch them. Hence, $\phi$ stays at $1$ as the direction of the predator, and the prey group's movement remains the same over time. 
 
\begin{figure}[!t]
	\centering  
\subfigure[ ]{\label{fig:6a}\includegraphics[width=6.0cm,height=5.5cm]{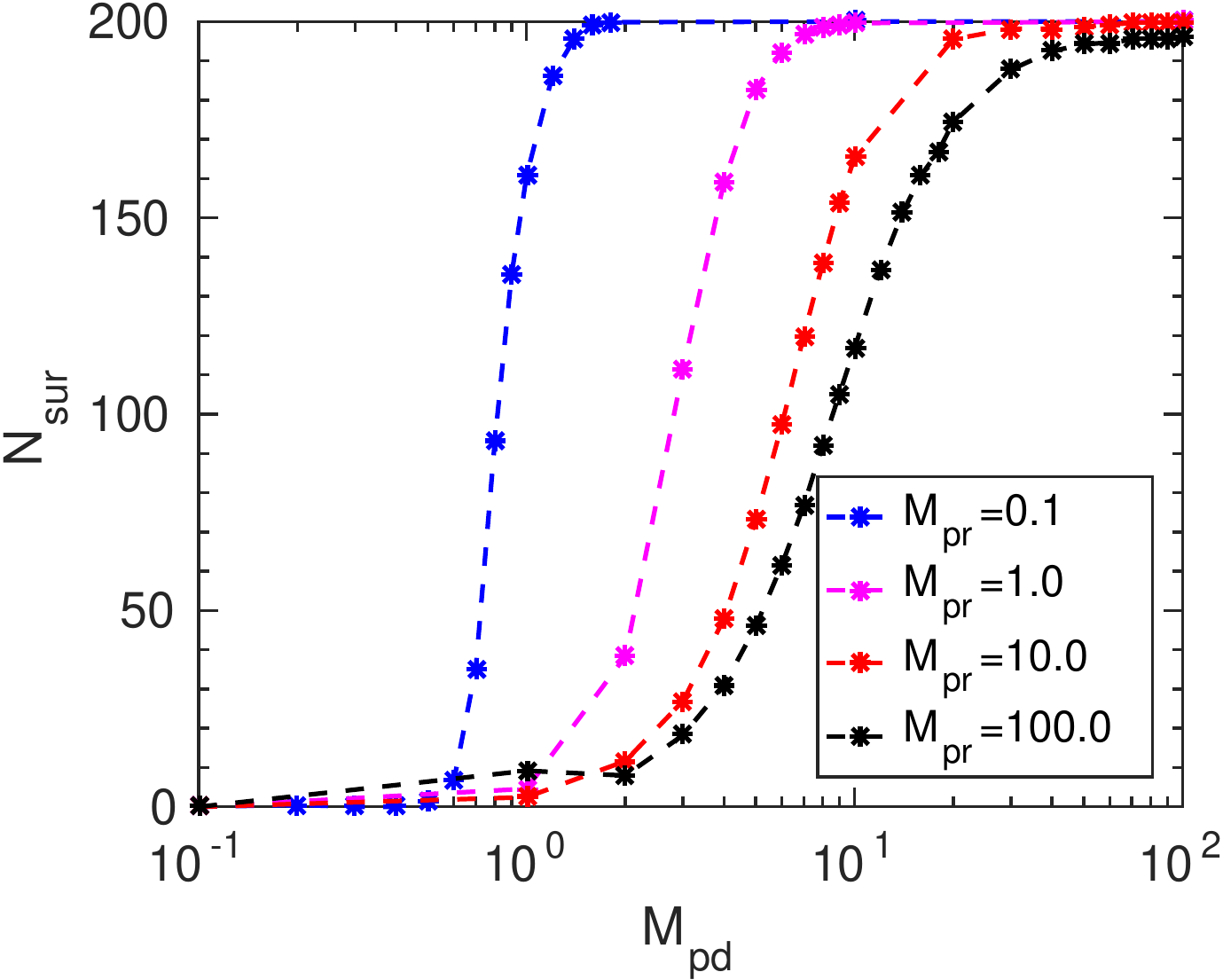}}	

\subfigure[ ]{\label{fig:6b}\includegraphics[width=7.0cm,height=6.5cm]{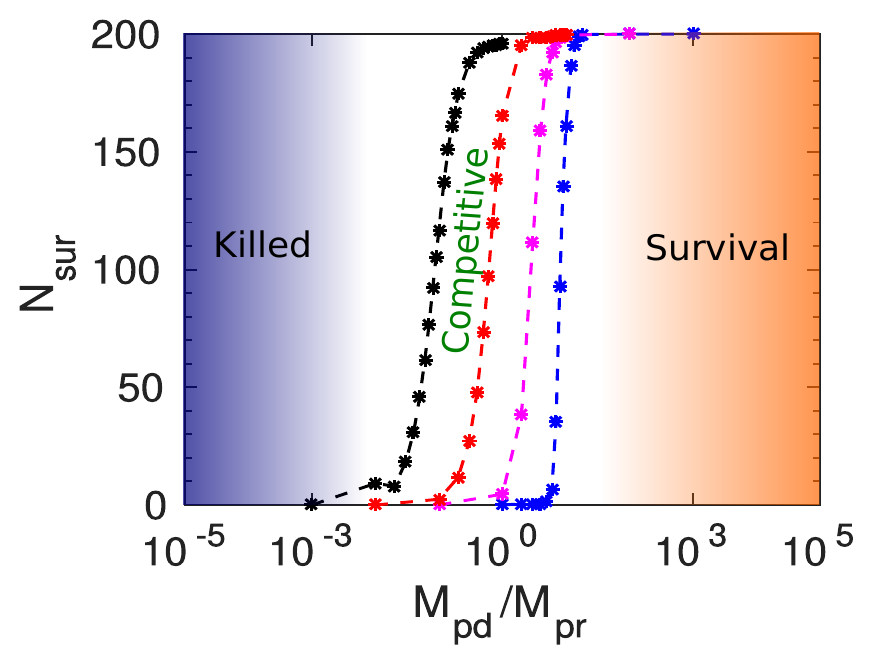}}

	\caption{(a) Number of survived prey, $N_{\rm sur}$, as a function of predator mass, $M_{\rm pd}$, for different values of prey mass, $M_{\rm pr}$, keeping the predator strength, $\delta_0= 2.5$, and initial prey group size, $N=200$. (b) The number of survived prey, $N_{\rm sur}$, has been plotted as a function of predator-to-prey mass ratio, $M_{\rm pd}/M_{\rm pr}$; three distinct regimes, {\it i.e.}, killed, competitive, and survival of prey swarm have been shown. }
	\label{fig:Nsur_vs_mpd}
\end{figure}

As seen, inertial forces significantly influence the escape trajectories, 
 we now ask the question: how the mass of the predator in comparison with the mass of the prey affects the survival of the prey group? In other words, whether a favourable ratio of $M_{\rm pd}/M_{\rm pr}$ exists for efficient predation in an ecosystem. 
 To answer the question, we now investigate the survival of the prey group with a varying mass of the predator and the prey.
We have simulated the dynamics for a sufficiently long time ($T=2000$) and plotted the number of survived prey, $N_{\rm sur}$, as a function of the mass of the predator, $M_{\rm pd}$, for different values of prey mass, $M_{\rm pr}$, where the predator strength has been kept constant (at a higher strength, $\delta_0=2.5$). 
Figure \ref{fig:Nsur_vs_mpd}(a) shows that 
the prey group is mostly killed at the lower mass regime of the predator.
Since the lighter predator can quickly change its moving direction due to the low inertia, the predator (with high predator strength) is capable of constant chasing, turning around, and capturing; thus, all prey are eventually killed (as presented in Figs. \ref{phase_plot}(a)-(d)). 
On the other hand, in the higher mass regime of the predator, 
most prey could survive.
 Higher predator mass, $M_{\rm pd}$, lowers the maneuverability of the predator due to the greater inertia. Thus, the heavier predator is unable to compete with the prey while chasing.  
However, in the intermediate mass regime, it exhibits a transition from non-survival to survival of the prey group (keeping the predator strength constant).
As seen from Fig. \ref{fig:Nsur_vs_mpd}(a), 
the transition regime also depends on the prey mass, $M_{\rm pr}$. For a constant predator mass, the survival of the lighter prey is more probable as their maneuverability is comparatively higher; thus, the prey groups divide, merge, and turn around the predator quickly and finally escape.

\begin{figure*}[!t]
	\centering  		
	\subfigure[ ]{\label{fig:6a}\includegraphics[width=5.5cm,height=5.5cm]{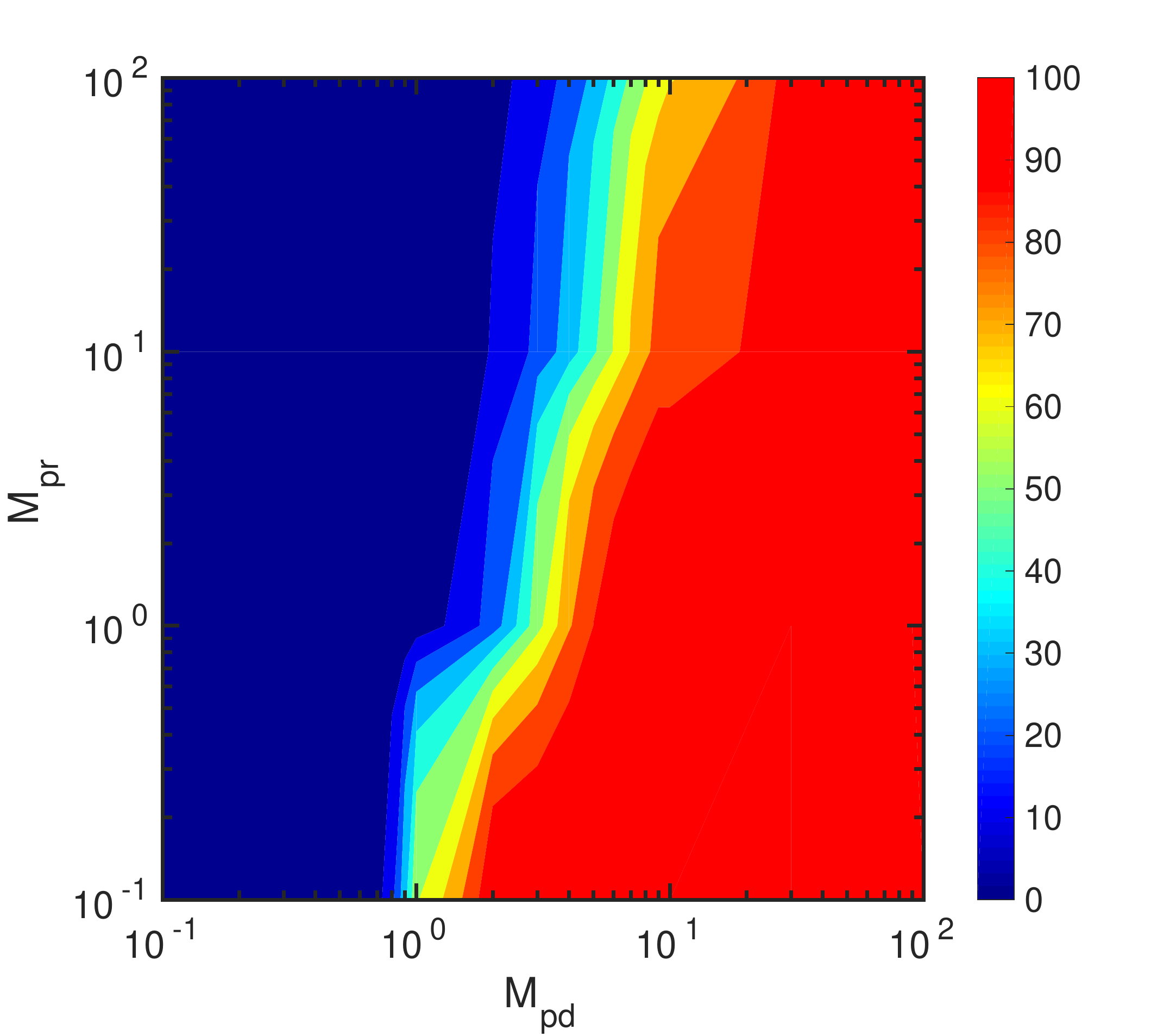}}
	\hspace{0.1cm}
	\subfigure[ ]{\label{fig:6b}\includegraphics[width=5.5cm,height=5.5cm]{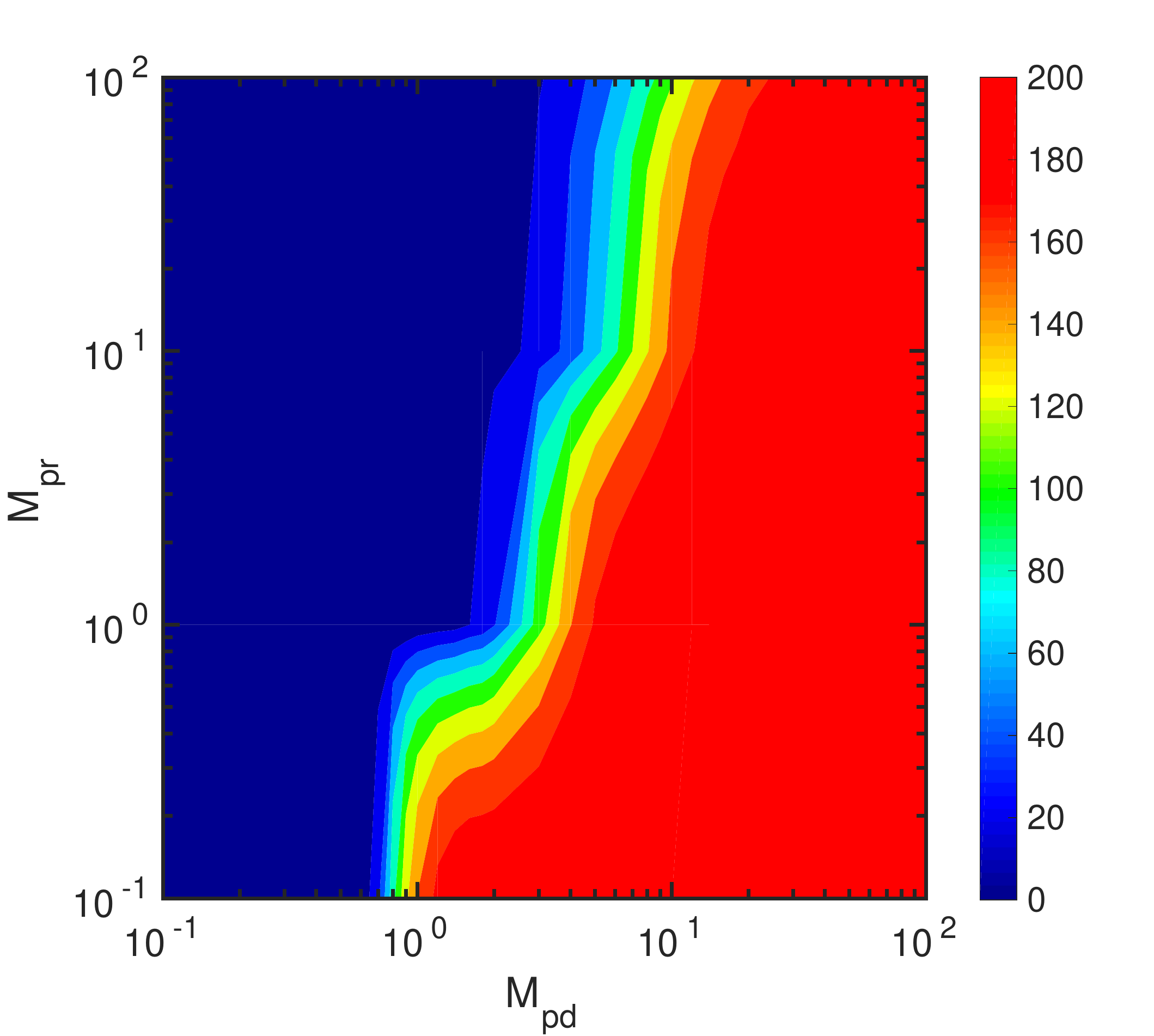}}
    \hspace{0.1cm}
	\subfigure[ ]{\label{fig:6b}\includegraphics[width=5.5cm,height=5.5cm]{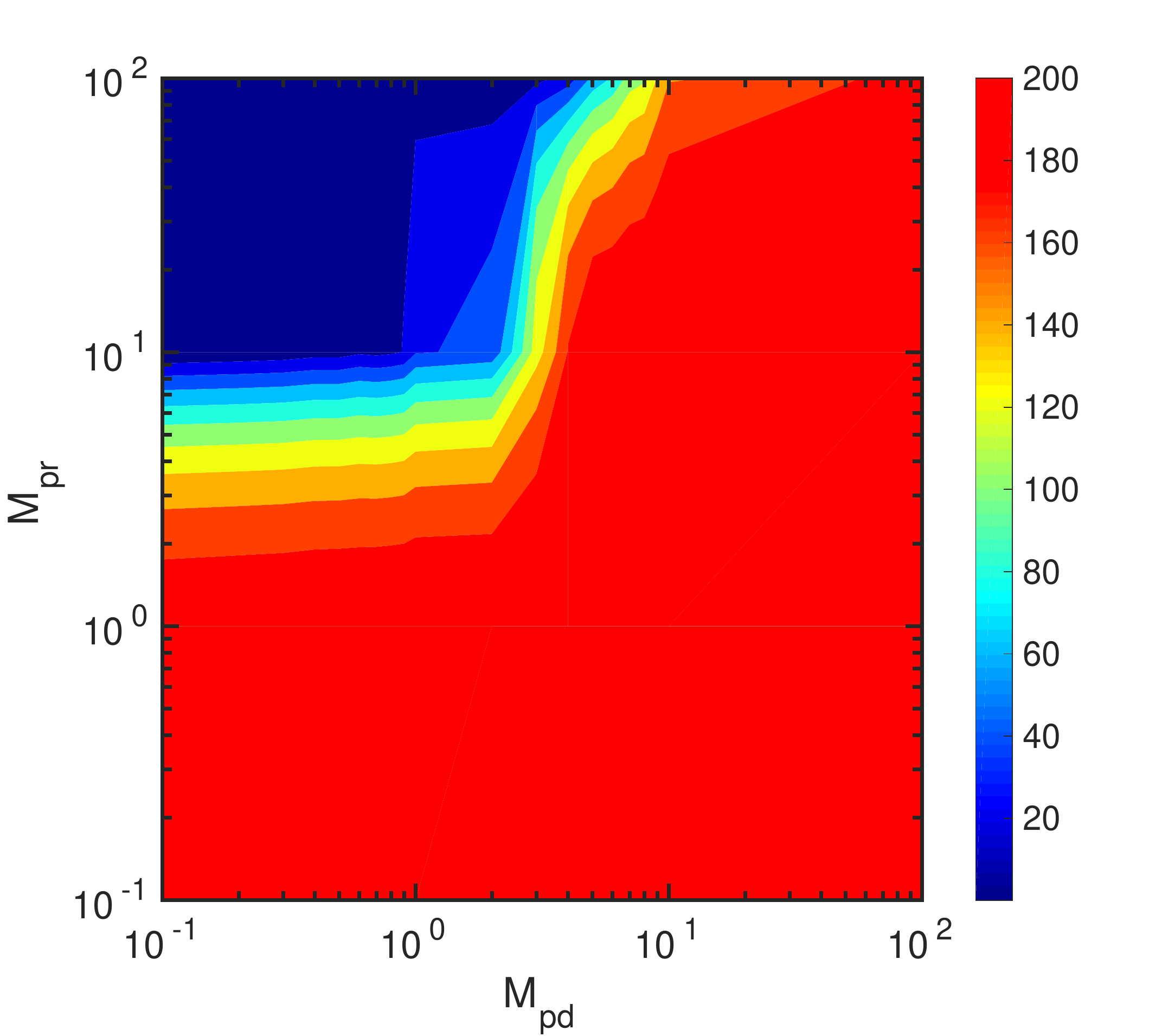}}
	
		\caption{Contour plots show the non-survival to survival regime of the prey swarm with the varying mass of the prey and the predator for two different initial prey group sizes, (a) $N=100$, (b) $N=200$, and predator strength, $\delta_0=2.5$; (c) shows the same for a weak predator, $\delta_0=0.4$ keeping prey group size, N=200. The colour bar represents the number of survived prey, $N_{\rm sur}$. (colour online)}
	\label{fig:contour_varyingN_varying_str}
\end{figure*}

Further, to quantify the transition regime, we have plotted the number of survived prey,  $N_{\rm sur}$, as a function of the ratio of predator-to-prey mass, $M_{\rm pd}/M_{\rm pr}$ in Fig. \ref{fig:Nsur_vs_mpd}(b).
It exhibits three distinct regimes specifying three different outcomes of survival and hunting of the prey group chased by the predator. 
The lower predator-prey mass ratio indicates the non-survival regime where all prey are captured. Interestingly, the intermediate predator-prey mass ratio unveils the competitive regime where prey and predator demonstrate an arms race for survival; here, some prey are survived, and some are killed. Whereas the higher predator-prey mass ratio shows the survival regime where all prey could defend against the predator attacks successfully. 
Many field studies have explored the dependence of the predator-to-prey mass ratio as a pivotal indicator for survival.
As predicted by our theoretical study, a favourable mass ratio is also observed for the hunting and survival of every prey-predator species in natural ecosystems \cite{woodward2007bookchapter,brose2010funeco}. 
Further, our study shows that the variation in initial prey group size does not affect the transition regime much if the predator strength is kept constant, as shown in Figs. \ref{fig:contour_varyingN_varying_str}(a)-(b). However, the strength of the predator has a profound effect on determining the survival regime. 
Figure \ref{fig:contour_varyingN_varying_str}(c)  shows the survival regime 
in the case of a weak predator, for $\delta_0=0.4$. 
Although the prey group could mostly survive, there exists a small parameter regime 
where the predator is able to catch some prey, 
in contrast to the extended parameter space of hunting in the case of a strong predator, $\delta_0=2.5$, as seen from Fig. \ref{fig:contour_varyingN_varying_str}(b).

\section*{Discussion}

Our study elucidates how inertial forces play an important role in determining the survival of a prey swarm under a predator attack.
Based on a simple theoretical model incorporating prey-prey and prey-predator interactions, we show how various escape trajectories emerge with varying the prey and predator mass and predator strength. 
In the case of a weak predator, we observe that the prey swarm forms a circular ring around the confused predator. However, as the predator mass increases, it exhibits a transition from stable ring formation to chasing dynamics. A detailed stability analysis calculations further validate this transition with varying mass.
Further, we have looked into the dynamics of a prey swarm when chased by a strong predator. With varying mass, we observed different escape patterns - prey swarm divides into subgroups, and merges again into a unitary group, turns into an arc to avoid the attack called F-maneuvering and escape, as seen in nature.
Our analysis shows that the attack frequency decreases with increasing predator mass as the maneuvering capability of the predator decreases with higher mass. 
Further, our study shows a transition from the non-survival to the survival of the prey swarm as the mass of the predator is increased, keeping the prey mass and predator strength fixed. 
Interestingly, as we plot the survived number of prey as a function of predator-to-prey mass ratio, it unveils three distinct regimes. The lower predator-prey mass ratio favours hunting; the intermediate-range shows competitive dynamics resembling a prey-predator arms race where some of the prey survive, and some are killed; and a higher predator-prey mass ratio which is most favourable for the survival of the prey swarm.
The theoretical findings of the existence of an optimal predator-to-prey mass ratio have also been observed in field studies of different prey-predator species in natural ecosystems \cite{woodward2007bookchapter,brose2010funeco}.
We have carried out further analysis with different initial prey group sizes and various predator strengths. Our study shows that the transition regime does not vary with the prey group size but is very much dependent on the predator strength.
Thus, our simple theoretical model could shed light on how inertial forces, together with the predator strength, could affect the pursuit and evasion dynamics of the prey swarm. 
This model could further be extended by incorporating empirical observations to get insights into the complexities of different prey-predator species in their natural habitat and also to study the collective motion of various living organisms.

\section*{ACKNOWLEDGEMENTS}

The authors acknowledge the financial support from Science and Engineering Research Board (SERB), Grant No. SR/FTP/PS-105/2013, Department of Science and Technology (DST), India. D.C acknowledges DST INSPIRE Fellowship for financial support.


\appendix
\renewcommand\thefigure{\thesection.\arabic{figure}}    
\section{Appendix-I}
\setcounter{figure}{0}

Here we analyze the prey-predator dynamics described by  Eq. \ref{preyequation} and Eq. \ref{predequation} in continuum forms. In the continuum limit, Eq. \ref{preyequation} can be written as
\begin{eqnarray}
M_{\rm pr}\frac{d\vec{v}(\vec{x},t)}{dt} + \vec{v}(\vec{x},t) &=& \int_{\mathbb{R}^2}[{\alpha_{0} \frac{\vec{x}-\vec{y}}{|\vec{x} -\vec{y}|^2} - \beta_{0}(\vec{x} - \vec{y})}] \rho(\vec{y},t) d\vec{y}\nonumber\\
&&  \;\;\;\;\;\;\;\;\;+\gamma_{0} \frac{\vec{x} - \vec{z}}{|\vec{x} - \vec{z}|^2},
\label{preyequation_cont}
\end{eqnarray}
Where, $\vec{R_{i}}$, $\vec{R_{j}}$, and $\vec{R_{p}}$ are written in continuum notations by $\vec{x}$, $\vec{y}$, and $\vec{z}$ respectively. $\vec{v}(\vec{x},t)=\frac{d\vec{x}}{dt}$ is the prey velocity at position $\vec{x}$ and time t. We use the fact that in the continuum limit, $\frac{1}{N_{sur}}\sum_{i=1}^{N_{sur}}\delta(\vec{x} - \vec{x_i})\approx \int_{\mathbb{R}^2}^{}\rho(\vec{y},t)d\vec{y}$ where
$\rho(\vec{y},t)$ is called the prey density distribution function which follows  $$\int_{\mathbb{R}^2}\rho_(\vec{y},t) d\vec{y}=1.$$ 
Similarly the equation of motion of the predator (Eq. \ref{predequation}) in the continuum limit can be written as
\begin{equation}
M_{\rm pd}\frac{d^2\vec{z}}{dt^2} + \frac{d\vec{z}}{dt}= \int_{\mathbb{R}^2}\frac{\vec{y} - \vec{z}}{|\vec{y} - \vec{z}|^p}\rho_(\vec{y},t) d\vec{y}.
\label{predequation_continuam}
\end{equation}
Now, the continuity equation is:

$$\frac{d\rho}{dt} + \nabla \cdot \rho(\vec{x},t)\vec{v}(\vec{x},t)=0.$$
Considering $\rho$ is spatially homogenous {\it i.e.} $\nabla \rho(\vec{x},t)=0$; Thus,
\begin{equation}
\nabla \cdot \vec{v}(\vec{x},t)= - \frac{1}{\rho} \frac{d\rho}{dt}.
\label{modified_continuity}
\end{equation}
Taking divergence on both sides of Eq. \ref{preyequation_cont} with respect to $\vec{x}$, we get
\begin{eqnarray}
M_{\rm pr}\Big[\nabla \cdot \frac{d\vec{v}}{dt}\Big] + \nabla \cdot \vec{v} = \int_{\mathbb{R}^2}\Big[\alpha_{0} \nabla \cdot \frac{\vec{x}-\vec{y}}{|\vec{x} -\vec{y}|^2} -\nonumber\\
 \beta_{0} \nabla \cdot (\vec{x} - \vec{y})\Big] \rho(\vec{y},t) d\vec{y} + \gamma_{0} \nabla \cdot \frac{\vec{x} - \vec{z}}{|\vec{x} - \vec{z}|^2}
\end{eqnarray}
Using the expression of ($\nabla \cdot \vec{v}$) from Eq. \ref{modified_continuity} in the above equation, we obtain the following equation:
\begin{eqnarray}
-M_{\rm pr}\frac{d}{dt}(\frac{1}{\rho} \frac{d\rho}{dt})- \frac{1}{\rho} \frac{d\rho}{dt}=\int_{\mathbb{R}^2}\Big[(2\pi\alpha_0 \delta(\vec{x}- \vec{y})-2\beta_0) \nonumber \\ \rho(\vec{y},t) d\vec{y}\Big] 
+ 2\pi\gamma_0\delta(\vec{x}-\vec{z}),
\label{prey_2}
\end{eqnarray}

where we have used the following identities:
$$\nabla \cdot \frac{\vec{x}-\vec{y}}{|\vec{x} -\vec{y}|^2}=2\pi\delta(\vec{x}- \vec{y}),$$
$$\nabla \cdot ({\vec{x}-\vec{y}})=2.$$
Since, $\delta(\vec{x}- \vec{z})=0$ (as $\vec{x} \neq \vec{z}$), we can write Eq. \ref{prey_2} as
$$-M_{\rm pr}\frac{d}{dt}(\frac{1}{\rho} \frac{d\rho}{dt})- \frac{1}{\rho} \frac{d\rho}{dt}=2\pi\alpha_0 \rho(\vec{x},t)-2\beta_0,$$

$$-M_{\rm pr}[\frac{1}{\rho} \frac{d^2\rho}{dt^2} - \frac{1}{\rho^2}(\frac{d\rho}{dt})^2] + \frac{1}{\rho} \frac{d\rho}{dt}=2\beta_0- 2\pi\alpha_0\rho(\vec{x},t).$$
Introducing a new variable $w=\frac{d\rho}{dt}$, the last equation becomes

\begin{equation}
-M_{\rm pr}[\frac{1}{\rho} \frac{dw}{dt} - \frac{1}{\rho^2}w^2] + \frac{w}{\rho}=2\beta_0- 2\pi\alpha_0 \rho(\vec{x},t).
\end{equation}
In the steady state, $\frac{d\rho}{dt}=0$ \& $\frac{dw}{dt}=0$. This gives $w^s=0$, and $\rho^s=\frac{\beta_0}{\pi\alpha_0}$, where $w^s$ and $\rho^s$ are the respective steady state values of $w$ and $\rho$. It turns out that $\rho^s$ is independent of predator and prey mass.

Now, in case of a weak predator, from the simulations, we find that in the steady state, the predator stays  at the centre and prey group circles around the predator, as shown in Fig. \ref{figure_2}a. Further, we analytically calculate the inner radius ($R_1$) and outer radius ($R_2$) of the annulus (A) formed by the prey group at the steady state, as shown in Fig. \ref{perturbed_configuration}. Thus we can write,

\begin{eqnarray}
M_{\rm pr}\frac{d\vec{v}}{dt} + \vec{v} &=& \int_{R_1}^{R_2}({\alpha_{0} \frac{\vec{x}-\vec{y}}{|\vec{x} -\vec{y}|^2} - \beta_{0}(\vec{x} - \vec{y})}) \rho(\vec{y},t) d\vec{y} \nonumber\\ && ~~~~~~~~~+ \gamma_{0} \frac{\vec{x} - \vec{z}}{|\vec{x} - \vec{z}|^2}.
\label{prey_3}
\end{eqnarray}
Using the identities $\int_{|\vec{y}|\le R}^{}\frac{\vec{x}-\vec{y}}{|\vec{x} -\vec{y}|^2}d\vec{y}=\pi \vec{x}$ for ${|\vec{x}|<R}$,
\newline
and $\int_{|\vec{y}|\le R}^{}\frac{\vec{x}-\vec{y}}{|\vec{x} -\vec{y}|^2}d\vec{y}=\pi R^2 \frac{\vec{x}}{|\vec{x}|.^2}$ for $|\vec{x}|>R$,
\newline
we get $\int_{0}^{R_2}\frac{\vec{x}-\vec{y}}{|\vec{x} -\vec{y}|^2}d\vec{y}=\pi \vec{x}$, and $\int_{0}^{R_1}\frac{\vec{x}-\vec{y}}{|\vec{x} -\vec{y}|^2}d\vec{y}=\pi R_1^2 \frac{\vec{x}}{|\vec{x}|.^2}$.
Considering polar symmetry, it can be shown that $\int_{R_1}^{R_2}\vec{y} d\vec{y}=0$.
Thus, Eq. \ref{prey_3} reduces to the following form:
 
\begin{eqnarray}
M_{\rm pr}\frac{d\vec{v}}{dt} + \vec{v} &=& \pi\rho\vec{x}[\alpha_0-\beta_0(R_2^2 - R_1^2)] - \pi\rho\alpha_0 R_1^2 \frac{\vec{x}}{|\vec{x}|^2} \nonumber \\ && ~~~~~~~ + \gamma_{0} \frac{\vec{x} - \vec{z}}{|\vec{x} - \vec{z}|^2},
\label{preyequation_continuam_1}
\end{eqnarray}
\newline
Using the relation ${\int_{0}^{R}\vec{y}d\vec{y}=\pi R^2}.$
At the steady state $\frac{d\vec{x}}{dt}=0$ and $\frac{d\vec{v}}{dt}=0$; let $\vec{x}=\vec{x^*}$ and $\vec{v}=\vec{v^*}$ be the respective steady state values of $\vec{x}$ and $\vec{v}$. Then, from Eq. \ref{preyequation_continuam_1}, we get

$$\pi\rho\vec{x^*} [\alpha_0 - \beta_0(R_2^2 - R_1^2)] + \gamma_{0} \frac{\vec{x^*} - \vec{z*}}{|\vec{x^*} - \vec{z^*}|^2} - \pi\rho\alpha_0 R_1^2 \frac{\vec{x}}{|\vec{x}|^2}=0,$$ where $\vec{z^*}$ is the steady state value of $\vec{z}$.
By assuming that the predator sits at the origin of the annulus at the steady state, we finally arrive at the following equation:
$$\pi\rho\vec{x^*} [\alpha_0 - \beta_0(R_2^2 - R_1^2)] + (\gamma_0 - \pi\rho\alpha_0 R_1^2)\frac{\vec{x^*}}{|\vec{x^*}|^2}=0.$$
For non-trivial solution,
$\alpha_0 -\beta_0(R_2^2 - R_1^2)=0$ and $\gamma_0 - \alpha_0\pi\rho R_1^2=0$. Hence,

$$R_1=\sqrt{\frac{\gamma_0}{\beta_0}},$$
$$R_2=\sqrt{\frac{\alpha_0 + \gamma_0}{\beta_0}}.$$
So, we can infer that the inner($R_{1}$) and outer($R_{2}$) radii of the ring depend on the prey-prey and prey-predator interaction strengths $\alpha_0$, $\beta_0$, $\gamma_0$ but are independent of prey and predator mass.


\appendix
\renewcommand\thefigure{\thesection.\arabic{figure}}    
\section{Appendix-II}
\setcounter{figure}{0}       

\begin{figure}[!t]
	\centering  		
	\includegraphics[width=5.5cm,height=5.5cm]{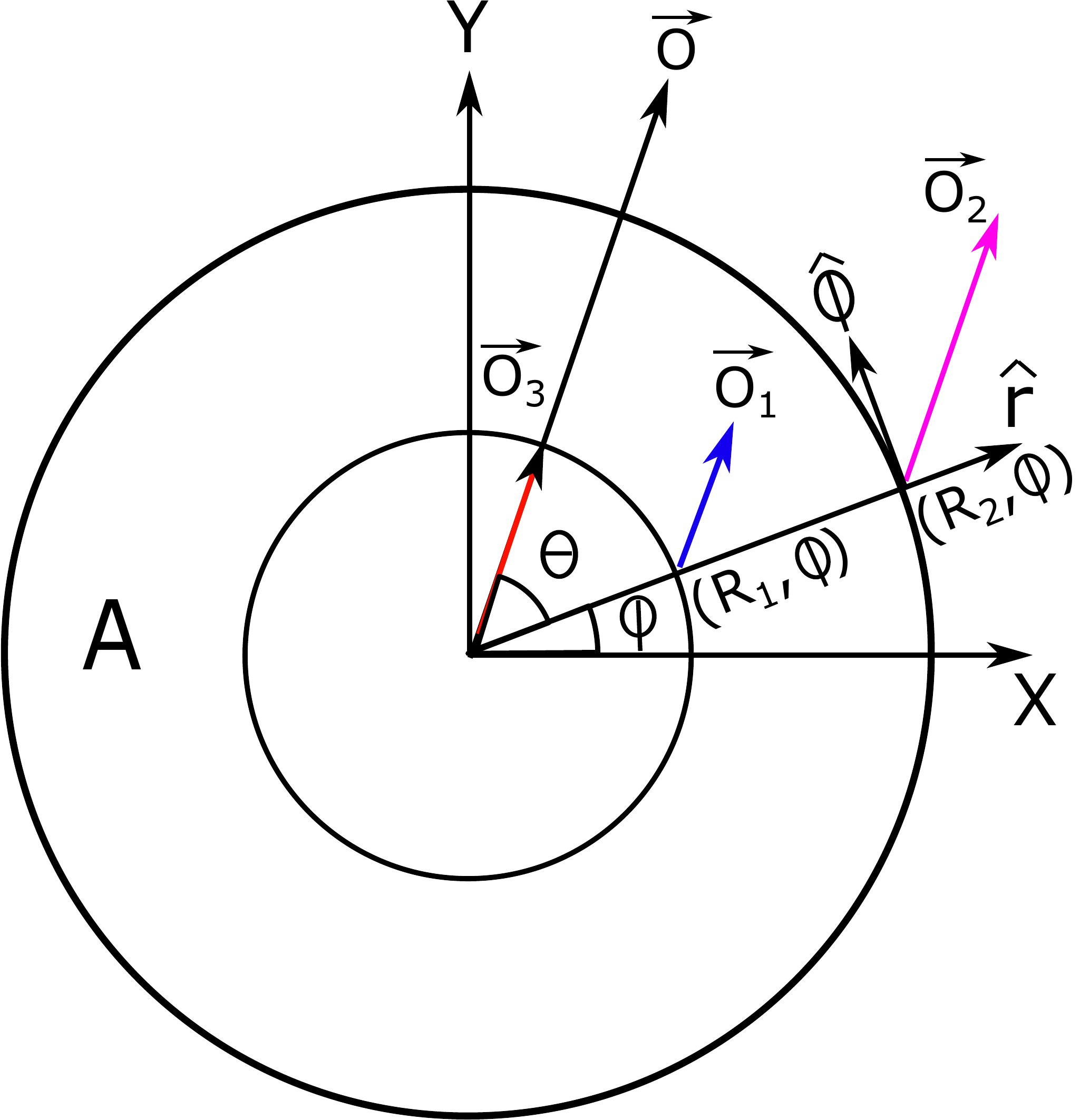}
	\caption{$\vec{O}$ is the general perturbation and $\vec{O_1}$, $\vec{O_2}$, $\vec{O_3}$ are its value at inner boundary, outer boundary and predator position. 'A' is specified as the annular region as shown in the diagram.}
	\label{perturbed_configuration}
\end{figure}

In this section, we analyze the stability criteria for the ring (A). For that we perturb the ring by a general perturbation vector $\vec{o}$ with an angle $\theta$ with respect to the origin and $\hat{r}$ as the direction of the perturbation, such that $\vec{o}=\vec{o_1}$ at the inner boundary, $\vec{o}=\vec{o_2}$ on the outer boundary and $\vec{o}=\vec{o_3}$ at the position of the predator as depicted in the Fig. \ref{perturbed_configuration}. The direction of the perturbation always stays constant along the direction $\hat{r}$. With perturbation $\vec{o}$, the Eq. \ref{prey_3} becomes

\begin{eqnarray}
M_{\rm pr}\frac{d\vec{v}}{dt} + \vec{v} &=& \rho \int_{R_1}^{R_2}\Big[{ \alpha_0 \frac{\vec{x}-(\vec{y}+\vec{o})}{|\vec{x} -(\vec{y}+\vec{o})|^2} - \beta_{0}(\vec{x} - (\vec{y} + \vec{o})}\Big]  d\vec{y} \nonumber \\ && ~~~~~~~~ + \gamma_{0} \frac{\vec{x} - \vec{o_3}}{|\vec{x} - \vec{o_3}|^2},
\label{preyequation_perturbed}
\end{eqnarray}

\begin{eqnarray*}
M_{\rm pr}\frac{d\vec{v}}{dt} + \vec{v} = \rho \int_{R_1}^{R_2}\Big[{ \alpha_0 \frac{\vec{u}-\vec{y}}{|\vec{u} -\vec{y}|^2} - \beta_{0}(\vec{u} - \vec{y})}\Big]  d\vec{y} + \nonumber \\ \gamma_{0} \frac{\vec{x} - \vec{o_3}}{|\vec{x} - \vec{o_3}|^2},
\end{eqnarray*}
where $\vec{u}=\vec{x}- \vec{o}$, such that $\vec{u}=\vec{u_2}=\vec{x}-\vec{o_2}$ on the outer boundary, and $\vec{u}=\vec{u_1}=\vec{x} - \vec{o_1}$ on the inner boundary.
Using the previous results $\int_{0}^{R_2}\frac{\vec{x}-\vec{y}}{|\vec{x} -\vec{y}|^2}d\vec{y}=\pi \vec{x}$, $\int_{0}^{R_1}\frac{\vec{x}-\vec{y}}{|\vec{x} -\vec{y}|^2}d\vec{y}=\pi R_1^2 \frac{\vec{x}}{|\vec{x}|.^2}$  and $\int_{R_1}^{R_2}\vec{y} d\vec{y}=0$, we get, 

\begin{eqnarray}
M_{\rm pr}\frac{d\vec{v}}{dt} + \vec{v} = \rho\Big[\pi\alpha_0 \vec{u_2} - \pi \alpha_0 R_{1}^2 \frac{\vec{u_1}}{|\vec{u_1}|^2} - \nonumber \\ \beta_0(\vec{u_2} \pi R_{2}^2 - \vec{u_1} \pi R_{1}^2)\Big] + \gamma_{0} \frac{\vec{x} - \vec{o_3}}{|\vec{x} - \vec{o_3}|^2}.
\end{eqnarray}
Substituting $\vec{u_1}$ and $\vec{u_2}$ in the above equation, we obtain
\begin{eqnarray}
M_{\rm pr}\frac{d\vec{v}}{dt} + \vec{v} = \rho\Big[(\pi \vec{x} -\pi \vec{o_2})\alpha_0 + \beta_0\pi \vec{x}(R_{1}^2 - R_{2}^2) \nonumber \\ - \pi \alpha_0 R_{1}^2 \frac{\vec{x} - \vec{o_1}}{|\vec{x} - \vec{o_1}|^2}  \beta_0 \pi(R_{2}^2 \vec{o_2} - R_{1}^2 \vec{o_1}\Big] + \gamma_{0} \frac{\vec{x} - \vec{o_3}}{|\vec{x} - \vec{o_3}|^2}.
\label{preyequation_perturbed_1}
\end{eqnarray}
At the steady state, $\vec{o}_{i}=0$, $\vec{v}=0$,$ \frac{d\vec{v}}{dt}=0$ and $\rho^s=\frac{\beta_0}{\alpha_0\pi}$ as stated above. Using these conditions in the Eq. \ref{preyequation_perturbed_1}, we get,
\begin{equation}
\pi \alpha_0 \rho + \pi \rho \beta_0(R_{1}^2 - R_{2}^2)=0.
\label{condition_1}
\end{equation}  
Combining Eq. \ref{preyequation_perturbed_1} and Eq. \ref{condition_1}, we finally arrive at the following equation:

\begin{eqnarray}
M_{\rm pr}\frac{d\vec{v}}{dt} + \vec{v} &=& \rho \pi[(\beta_0 R_2^2-\alpha_0)\vec{o_2} - \beta_0 R_1^2 \vec{o_1}] + \rho \pi \alpha_0 R_1^2[\frac{\vec{x}}{|\vec{x}|^2} \nonumber \\
&& ~ - \frac{\vec{x} - \vec{o_1}}{|\vec{x}- \vec{o_1}|^2}] + \gamma_{0} [\frac{\vec{x} - \vec{o_3}}{|\vec{x} - \vec{o_3}|^2} - \frac{\vec{x}}{|\vec{x}|^2}].
\label{preyequation_perturbed_2}
\end{eqnarray}
Next, we linearize the Eq. \ref{preyequation_perturbed_2} on the inner boundary, outer boundary and at the predator position. On the inner boundary, $$\vec{x}=R_1 \hat{r} + \vec{o_1}(\hat{r},\hat{\theta}).$$ \\
Hence, on the inner boundary, we can show

\begin{equation}
\frac{\vec{x}}{|\vec{x}|^2} -  \frac{\vec{x} - \vec{o_1}}{|\vec{x}- \vec{o_1}|^2}= \frac{1}{R_1^2}[\vec{o_1} - 2o_1cos\theta\hat{r}],
\label{eq:22}
\end{equation}
and 
\begin{equation}
\frac{\vec{x} - \vec{o_3}}{|\vec{x}- \vec{o_3}|^2} - \frac{\vec{x}}{|\vec{x}|^2} = -\frac{1}{R_1^2}[\vec{o_3} - 2o_3cos\theta\hat{r}],
\label{eq:23}
\end{equation}
where the terms other than the linear order are neglected due to smallness. Using Eq. \ref{eq:22} and Eq. \ref{eq:23}, the Eq. \ref{preyequation_perturbed_2} can be written as:
\begin{eqnarray}
M_{\rm pr}\frac{d\vec{v}}{dt} + \vec{v} &=& \rho \pi((\beta_0 R_2^2 - \alpha_0)\vec{o_2} - \beta_0 R_1^2 \vec{o_1}) + \rho \pi \alpha_0(\vec{o_1} \nonumber \\
&& ~ - 2o_1cos\theta\hat{r}) - \frac{\gamma_0}{R_1^2}[\vec{o_3} - 2o_3cos\theta\hat{r}].
\label{eq:prey_perturbed3}
\end{eqnarray}
Since, $\hat{r}$ is taken as a constant, $\vec{v}=\frac{d\vec{o_1}}{dt}$ and $\frac{d\vec{v}}{dt}= \frac{d^2\vec{o_1}}{dt^2}$ on the inner boundary. Hence, the Eq. \ref{eq:prey_perturbed3} becomes

\begin{eqnarray}
M_{\rm pr}\frac{d^2\vec{o_1}}{dt^2} + \frac{d\vec{o_1}}{dt} =\rho \pi((\beta_0 R_2^2 - \alpha_0)\vec{o_2} - \beta_0 R_1^2 \vec{o_1}) + \nonumber \\ \rho \pi\alpha_0(\vec{o_1}- 2o_1cos\theta\hat{r}) - \frac{\gamma_0}{R_1^2}[\vec{o_3} - 2o_3cos\theta\hat{r}].
\label{eq:prey_perturbed3_1}
\end{eqnarray}
Taking scalar product on both sides of Eq. \ref{eq:prey_perturbed3_1} by $\hat{r}$, the Eq. \ref{eq:prey_perturbed3_1} reduces to the following form:
\begin{equation}
M_{\rm pr}\frac{d^2 o_1}{dt^2} + \frac{d o_1}{dt}= \rho \pi(-\beta_0 R_1^2 - \alpha_0)o_1 + \rho \pi(\beta_0 R_2^2 - \alpha_0)o_2 + \frac{\gamma_0 o_3}{R_1^2}.
\label{eq:prey_preturbed_dot_product}  
\end{equation}
In terms of the system parameters, the last equation is
\begin{equation}
M_{\rm pr}\frac{d^2 o_1}{dt^2} + \frac{d o_1}{dt}=-\frac{\beta_0}{\alpha_0}(\gamma_0 + \alpha_0) o_1 +\frac{\beta_0 \gamma_0}{\alpha_0} o_2 + \beta_0 o_3.
\label{o1}
\end{equation}
Next, we linearize the Eq. \ref{preyequation_perturbed_2} on the outer boundary where $\vec{x}=R_2 \hat{r} + \vec{o_2}$. Using similar calculations as done previously, we get the following equations:

\begin{eqnarray}
\frac{\vec{x}}{|\vec{x}|^2} -  \frac{\vec{x} - \vec{o_1}}{|\vec{x}- \vec{o_1}|^2}= \frac{1}{R_2^2}[\vec{o_1} - 2o_1cos\theta\hat{r}], \\
\frac{\vec{x} - \vec{o_3}}{|\vec{x}- \vec{o_3}|^2} - \frac{\vec{x}}{|\vec{x}|^2} = -\frac{1}{R_2^2}[\vec{o_3} - 2o_3cos\theta\hat{r}], \\
M_{\rm pr}\frac{d^2 o_2}{dt^2} + \frac{d o_2}{dt}=-\frac{\beta_0 \gamma_0}{\alpha_0}(1 + \frac{\alpha_0}{\alpha_0 + \gamma_0})o_1 + \nonumber \\ \frac{\beta_0 \gamma_0}{\alpha_0} o_2 +  \frac{\beta_0 \gamma_0}{\alpha_0 + \gamma_0}o_3.
\label{o2}
\end{eqnarray}
We next linearize the predator equation (Eq. \ref{predequation_continuam}). Since, $\vec{o_3}$ is the perturbation at the position of the predator about its steady state(origin), therefore $\vec{z}=\vec{o_3}$.
\newline
Hence, with perturbation, the Eq. \ref{predequation_continuam} becomes

\begin{equation}
M_{\rm pd}\frac{d^2\vec{o_3}}{dt^2} + \frac{d\vec{o_3}}{dt}=\delta_0 \rho \int_{B(o_2,R_2)/(o_1,R_1)} \frac{\vec{x} - \vec{o_3}}{|\vec{x} - \vec{o_3}|^p} d\vec{x}, 
\label{predator_equation}
\end{equation}
where $B(o_{2},R_{2})/(o_{1},R_{1})$ denotes the annular region A with perturbations $o_{1}$ and $o_{2}$ on its inner and outer boundaries respectively. Now, $B(o_{2},R_{2})/(o_{1},R_{1})$ and A follows the following transformation relation:
\begin{equation}
\int_{B(o_2,R_2)/(o_1,R_1)} \frac{\vec{x}}{x^p} d\vec{x}=\int_{A}\frac{\vec{x} + \vec{o}}{|\vec{x} + \vec{o}|^p} d\vec{x}.
\label{Identity}
\end{equation}
Assuming $|\vec{o_3}|^2$ is small, we can show that $$\frac{\vec{x} - \vec{o_3}}{|\vec{x} - \vec{o_3}|^2}\approx \frac{\vec{x}}{x^p} + \frac{p\vec{x} (\vec{x}\cdot \vec{o_3}) - \vec{o_3}|\vec{x}|^2}{x^{p+2}}.$$
So, the Eq. \ref{predator_equation} becomes
\begin{eqnarray}
M_{\rm pd}\frac{d^2\vec{o_3}}{dt^2} + \frac{d\vec{o_3}}{dt}\approx \delta_0 \rho \int_{B(o_2,R_2)/(o_1,R_1)} \\ \nonumber [\frac{\vec{x}}{x^p} + \frac{p\vec{x} (\vec{x}\cdot \vec{o_3}) - \vec{o_3}|\vec{x}|^2}{x^{p+2}}] d\vec{x}.
\label{pre}
\end{eqnarray}
Using the transformation relation Eq. \ref{Identity}, one can can show that
\begin{equation}
\int_{B(o_2,R_2)/(o_1,R_1)} \frac{\vec{x}}{x^p} d\vec{x}\approx \int_{A} \frac{\vec{x}}{x^p} d\vec{x} + \vec{o} \int_{A} \frac{d\vec{x}}{x^p} - p\int_{A}\frac{\vec{x}(\vec{x} \cdot \vec{o})}{x^{p+2}}.
\label{predator_eq_2}
\end{equation}
Using $\vec{x}=x[\hat{i} \cos\theta + \hat{j} \sin\theta]$, and $d\vec{x}=xdxd\theta$, we can find the following integration results:

$$\int_{A} \frac{\vec{x}}{x^p} d\vec{x}=0,$$
$$\int_{A} \frac{\vec{o}}{x^p}d\vec{x}=\frac{2\pi}{2-p}[\vec{o_2}R_2^{2-p} - \vec{o_1}R_1^{2-p}],$$
$$\int_{A}\frac{\vec{x}(\vec{x} \cdot \vec{o})}{x^{p+2}} d\vec{x}= \frac{\pi R_2^{2-p} \vec{o_2} - \pi R_1^{2-p} \vec{o_1}}{2-p}.$$

So, the Eq. \ref{predator_eq_2} becomes
\begin{eqnarray}
\int_{B(o_2,R_2)/(o_1,R_1)} \frac{\vec{x}}{x^p} d\vec{x} &\approx& \pi(R_{2}^{2-p}\vec{o_{2}}-R_{1}^{2-p}\vec{o_{1}}).
\end{eqnarray}

Next, we need to integrate the following integration:
\begin{equation}
\int_{B(o_2,R_2)/(o_1,R_1)} \frac{p\vec{x} (\vec{x}\cdot \vec{o_3}) - \vec{o_3}|\vec{x}|^2}{x^{p+2}} d\vec{x}$$
$$=p \int_{B(o_2,R_2)/(o_1,R_1)} \frac{\vec{x}(\vec{x} \cdot \vec{o_3})}{|\vec{x}|^{p+2}} -\vec{o_3} \int_{B(o_2,R_2)/(o_1,R_1)} \frac{d\vec{x}}{x^p}
\label{predator_eq_3}
\end{equation}
$=I_{1}+I_{2}$,
\newline
where $I_{1}=p \int_{B(o_2,R_2)/(o_1,R_1)} \frac{\vec{x}(\vec{x} \cdot \vec{o_3})}{|\vec{x}|^{p+2}}$, 
\newline
and
\newline
$I_{2}=-\vec{o_3} \int_{B(o_2,R_2)/(o_1,R_1)} \frac{d\vec{x}}{x^p}$.
\newline
Using the transformation relation Eq. \ref{Identity}, we can write  $$I_{1}=p \int_A \frac{(\vec{x} + \vec{o})((\vec{x} + \vec{o}) \cdot \vec{o_3})}{|\vec{x} + \vec{o}|^{p+2}}.$$
Considering only the linear order terms in $\vec{o}$ and $\vec{o_{3}}$, we can show  
\begin{equation}
I_{1}= \frac{\pi p \vec{o_3}}{2-p} (R_2^{2-p} - R_1^{2-p}). 
\label{predator_eq_4}
\end{equation}
Similarly, we can show
\begin{equation}
I_{2}=-\frac{2\pi\vec{o_3}}{2-p}(R_2^{2-p} - R_1^{2-p}).
\label{predator_eq_5}
\end{equation}

Putting the expressions of $I_{1}$ and $I_{2}$ from Eq. \ref{predator_eq_4} and Eq. \ref{predator_eq_5}, finally Eq. \ref{predator_eq_3} becomes

\begin{equation}
\int_{B(o_2,R_2)/(o_1,R_1)} \frac{p\vec{x} (\vec{x}\cdot \vec{o_3}) - \vec{o_3}|\vec{x}|^2}{x^{p+2}} d\vec{x} = -\pi \vec{o_3}(R_2^{2-p} - R_1^{2-p}).
\label{predator_eq_6}
\end{equation}

Now the Eq. \ref{pre} can be written as:
\begin{eqnarray*}
M_{\rm pd}\frac{d^2\vec{o_3}}{dt^2} + \frac{d\vec{o_3}}{dt}\approx\rho \delta_0[-\pi R_1^{2-p} \vec{o_1} + \pi R_2^{2-p} \vec{o_2} +\\ \nonumber \pi \vec{o_3}(R_1^{2-p} - R_2^{2-p})]. 
\end{eqnarray*}
In terms of the system parameters, the last equation becomes

\begin{eqnarray}
M_{\rm pd}\frac{d^2\vec{o_3}}{dt^2} + \frac{d\vec{o_3}}{dt} &=& \frac{\delta_0 \beta_0}{\alpha_0}\Big[- (\frac{\gamma_0}{\beta_0})^{\frac{2-p}{2}} \vec{o_1} + (\frac{\alpha_0 + \gamma_0}{\beta_0})^{\frac{2-p}{2}} \vec{o_2} \nonumber \\ && ~~ + [\frac{\gamma_0}{\beta_0})^{\frac{2-p}{2}} - (\frac{\alpha_0+\gamma_0}{\beta_0})^{\frac{2-p}{2}}] \vec{o_3}\Big].
\end{eqnarray}

Since $\vec{o_{1}}$, $\vec{o_{2}}$, $\vec{o_{3}}$ are alligned along the same direction, therefore, linerized predator equation will be
\begin{eqnarray}
M_{\rm pd}\frac{d^2 o_3}{dt^2} + \frac{d o_3}{dt} &=& \frac{\delta_0 \beta_0}{\alpha_0} \Big[- (\frac{\gamma_0}{\beta_0})^{\frac{2-p}{2}} o_1 + (\frac{\alpha_0 +\gamma_0}{\beta_0})^{\frac{2-p}{2}}o_2 \nonumber \\ && ~ +[(\frac{\gamma_0}{\beta_0})^{\frac{2-p}{2}} - (\frac{\alpha_0 + \gamma_0}{\beta_0})^{\frac{2-p}{2}}]o_3\Big].
\label{predator_equation_final}
\label{o3}
\end{eqnarray}

\begin{figure*}[!t]
	\subfigure[ ]{\label{fig:mpr_1pt0}\includegraphics[width=6.0cm,height=5.5cm]{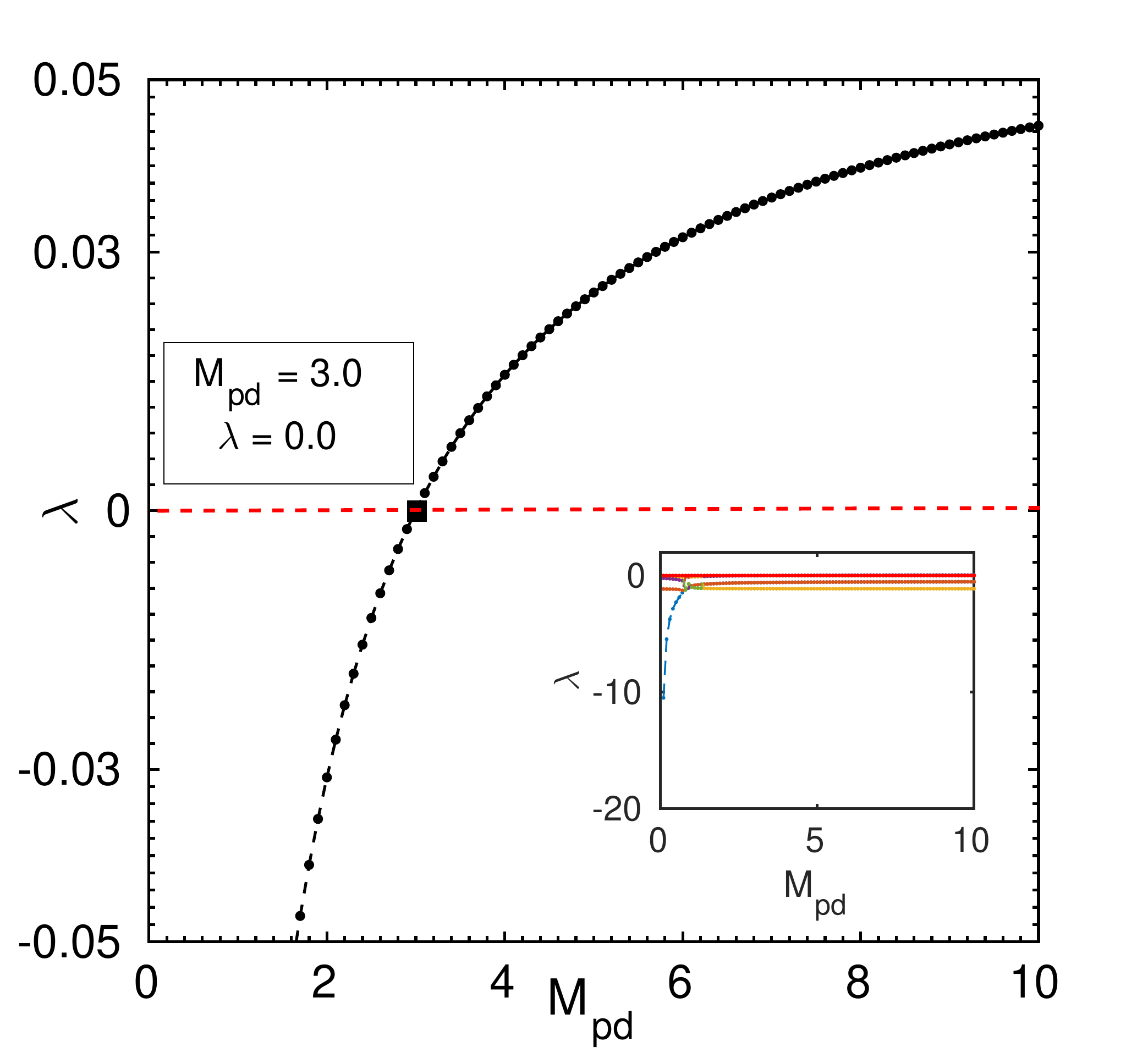}}
	\hspace{1.0cm}
   \subfigure[ ]{\label{fig:mpr_0pt1}\includegraphics[width=6.0cm,height=5.5cm]{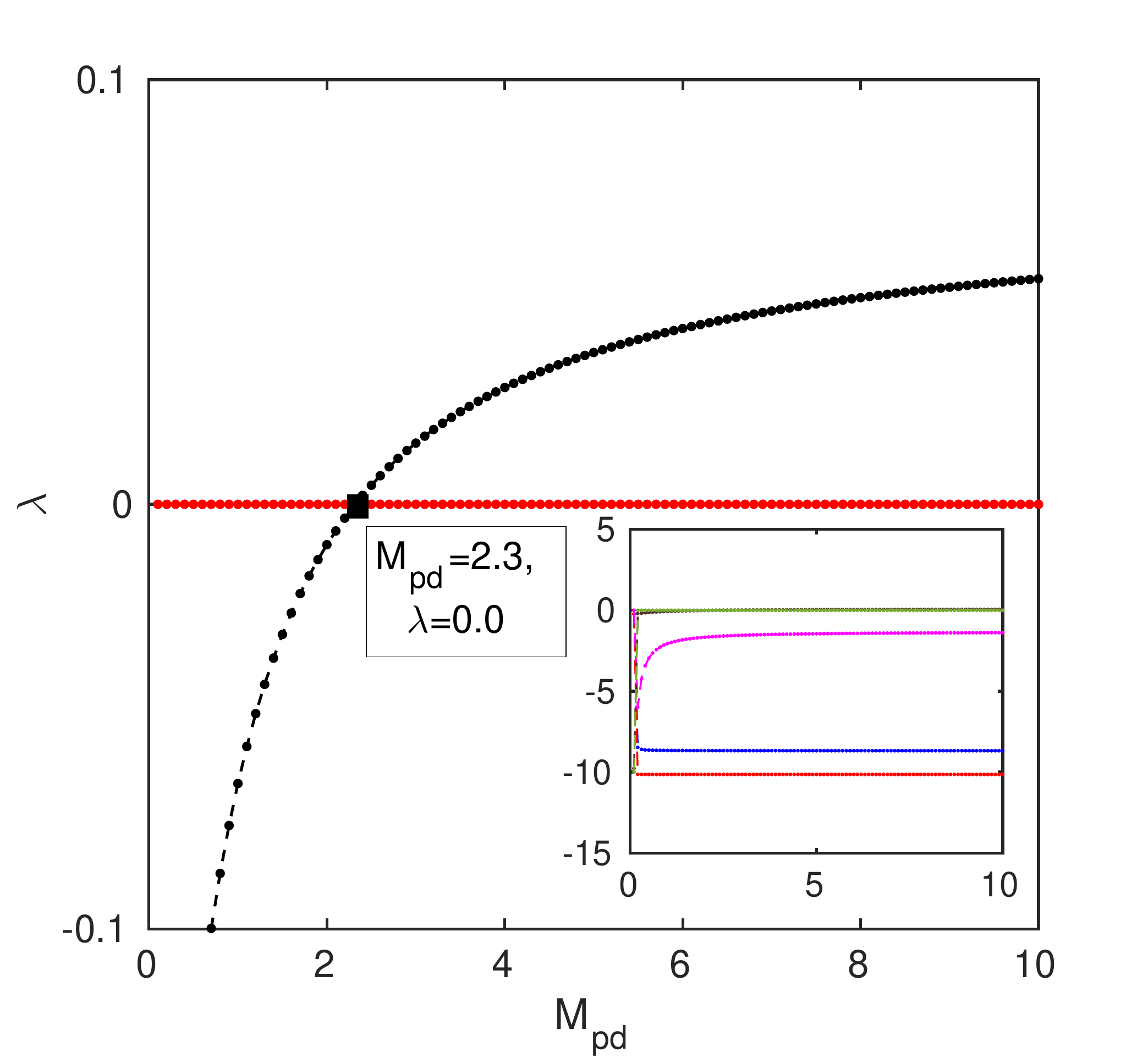}}
 	\caption{Eigenvalue has been plotted as a function of predator mass $M_{\rm pd}$ in (a) and (b) where the prey mass $M_{\rm pr}$ is kept constant to 1.0 and 0.1 respectively. Eigenvalue crosses zero at $M_{\rm pd}=3.0$ and $M_{\rm pd}=2.3$ in case of (a) and (b) respectively. Other eigenvalues of the transition matrix are negative as shown in the insets.}
 	\end{figure*}

Introducing a set of new variables $\vec{u}_{i}=\frac{d\vec{o}_{i}}{dt}$, where $i=1,2,3$. Finally, we get the Jacobian matrix as shown in \ref{matrix_equation}.

Eq. \ref{o1}, Eq. \ref{o2} and Eq. \ref{o3} can be written in matrix form as shown in Eq. \ref{matrix_equation}. The dynamics of the system could be predicated by analyzing the eigenvalues of the transition matrix $M$. Numerically, the steady state structure has been observed at $\delta_0=0.4, \beta_0=1.0, \gamma_0=0.2, \alpha_0=1.0$, $p=3$ for smaller prey and predator masses. Hence, we put the similar parameters in the Jacobian matrix $M$ and then investigate how the eigenvalues behave as we vary the prey and predator masses. Fig. \ref{fig:mpr_1pt0} clearly demonstrates the evolution of eigenvalues as we change the $M_{\rm pd}$; one of the eigenvalues goes from positive to negative at $M_{\rm pd}=3.0$ whereas other eigenvalues remain negative. Further, we have also investigated the eigenvalues for another mass of prey, $M_{\rm pr}=0.1$, as depicted in Fig. \ref{fig:mpr_0pt1}; here also one of the eigenvalues show the transition from negative to positive values at a predator mass, $M_{\rm pd}=2.3$. Thus, it is evident from the eigenvalue analysis that the stable ring becomes unstable at a particular predator mass, keeping the prey mass constant. Hence, we could infer that inertia of both prey and the predator have a profound effect on the prey-predator dynamics.

\begin{widetext} 
\begin{equation}
M=\begin{pmatrix}
\frac{d\vec{o_1}}{dt} \\ \frac{d\vec{o_2}}{dt} \\ \frac{d\vec{o_3}}{dt} \\ \frac{d\vec{u_1}}{dt} \\ \frac{d\vec{u_2}}{dt} \\ \frac{d\vec{u_3}}{dt}\\
\end{pmatrix} = \begin{pmatrix}
0 & 0 & 0 & 1 & 0 & 0 \\ 0 & 0 & 0 & 0 & 1 & 0 \\ 0 & 0 & 0 & 0 & 0 & 1 \\ -\frac{\beta_0}{\alpha_0 M_{\rm pr}}(\gamma_0 + \alpha_0) & \frac{\beta_0 \gamma_0}{\alpha_0 M_{\rm pr}} & \frac{\beta_0}{M_{\rm pr}} & -\frac{1}{M_{\rm pr}} & 0 & 0 \\ -\frac{\beta_0 \gamma_0}{\alpha_0 M_{\rm pr}}(1 + \frac{\alpha_0}{\alpha_0 + \gamma_0}) & \frac{\beta_0 \gamma_0}{\alpha_0 M_{\rm pr}} & \frac{\beta_0 \gamma_0}{M_{\rm pr}(\alpha_0 + \gamma_0)} & 0 & -\frac{1}{M_{\rm pr}} & 0 \\
-\frac{\delta_0 \beta_0}{\alpha_0 M_{\rm pd}} (\frac{\gamma_0}{\beta_0})^\frac{2-p}{2} & \frac{\delta_0 \beta_0}{\alpha_0 M_{\rm pd}} (\frac{\alpha_0 + \gamma_0}{\beta_0})^\frac{2-p}{2} & \frac{\delta_0 \beta_0}{\alpha_0 M_{\rm pd}}[(\frac{\gamma_0}{\beta_0})^\frac{2-p}{2} - (\frac{\alpha_0 + \gamma_0}{\beta_0})^\frac{2-p}{2}] & 0 & 0 & -\frac{1}{M_{\rm pd}}
\end{pmatrix} \begin{pmatrix}
\vec{o_1} \\ \vec{o_2} \\ \vec{o_3} \\ \vec{u_1} \\ \vec{u_2} \\ \vec{u_3} ,
\end{pmatrix}
\label{matrix_equation}
\end{equation}
\end{widetext}



%


\begin{thebibliography}{40}%

\bibitem{Kparrishscience}Parrish JK, Edelstein-Keshet L, Science. \textbf{284}, 5411 (1999).
\bibitem{sumpterbook2010}Sumpter DJ, Collective animal behavior. Princeton University Press.(2010)

\bibitem{vijay2020}Kumar, V., \& De, R, Royal Society open science. \textbf{8}, 9 (2021)
\bibitem{krausebook2002}Krause J, Ruxton GD, Living in groups. Oxford University Press.(2002)
\bibitem{ballerini2008}Ballerini M, Cabibbo N, Candelier R, Cavagna A, Cisbani E, Giardina I, Orlandi A, Parisi G, Procaccini A, Viale M, Zdravkovic V, Animal behaviour. \textbf{76},1 (2008)

\bibitem{Partho2019}De, P. S., \& De, R, Physical Review E. \textbf{100(1)}, 012409 (2019)

\bibitem{Arnoldplosone2012}Waters A, Blanchette F, Kim AD, PLoS One. \textbf{7}, 11 (2012) 
\bibitem{jmichaelscience2009}Makris NC, Ratilal P, Jagannathan S, Gong Z, Andrews M, Bertsatos I, Godø OR, Nero RW, Jech JM, Science. \textbf{323}, 5922 (2009)

\bibitem{debangana2019}Mukhopadhyay, D., \& De, R, Physical biology. \textbf{16(4)}, 046006.(2019)

\bibitem{debangana2022}Mukhopadhyay, D., \& De, R, Biophysical Journal. {\bf 121(3)} (2022)
	
\bibitem{traniello2003ARE}Traniello JF, Annual review of entomology. \textbf{34(1)}, (1989)

\bibitem{Dipanjan-Review2022}De, R., \& Chakraborty, D,  Journal of Biosciences. {\bf 47(3)} (2022)

\bibitem{penzhornbook1984}Penzhorn BL, Zeitschrift für Tierpsychologie. \textbf{64(2)}, (1984)

\bibitem{pitcherbook1983}Pitcher TJ, Wyche CJ, Predator-avoidance behaviours of sand-eel schools: why schools seldom split. InPredators and prey in fishes. Springer Dordrecht.(1983)

\bibitem{neilletal}Neill S, Cullen JM, Journal of Zoology. \textbf{172(4)}, (1974)
\bibitem{parrishenvbio}Parrish JK, Environmental Biology of Fishes. \textbf{55(1-2)} (1999)

\bibitem{Oshaninproceedings2009}Oshanin G, Vasilyev O, Krapivsky PL, Klafter J, Proceedings of the National Academy of Sciences. \textbf{106(33)} (2009).

\bibitem{Siddharth2020}Patwardhan, S., De, R., \& Panigrahi, P. K, The European Physical Journal E. { \bf 43(8)}  (2020) 

\bibitem{vicsekprl1995}Vicsek, T., Czirók, A., Ben-Jacob, E., Cohen, I. and Shochet, O., Physical review letters. \textbf{75(6)} (1995).

\bibitem{AngelaniPRL2012}Angelani L. Physical review letters. {\bf 109(11)} (2012).

\bibitem{Lettetal2014theoeco}Lett, C., Semeria, M., Thiebault, A. and Tremblay, Y., Theoretical Ecology.\textbf{7(3)} (2014)


\bibitem{Chen2014} Chen, Y., \& Kolokolnikov, T,  Journal of The Royal Society Interface. { \bf 11(94)}(2014) 

\bibitem{chakraborty2020}Chakraborty, D., Bhunia, S., \& De, R, Scientific Reports. \textbf{10(1)}(2020).

\bibitem{wilson2018nature}Wilson, A.M., Hubel, T.Y., Wilshin, S.D., Lowe, J.C., Lorenc, M., Dewhirst, O.P., Bartlam-Brooks, H.L., Diack, R., Bennitt, E., Golabek, K.A. and Woledge, R.C., Nature, \textbf{554(7691)} (2018)

\bibitem{barnes2010ecology}Barnes, C., Maxwell, D., Reuman, D.C. and Jennings, S., Ecology. \textbf{91(1)} (2010)

\bibitem{woodward2007bookchapter}Woodward, G. and Warren, P.H., Body size and predatory interactions in freshwaters: scaling from individuals to communities. Body size: the structure and function of aquatic ecosystems, pp.98-117,(2007)

\bibitem{brose2010funeco}Brose, U., Functional Ecology. \textbf{24(1)} (2010)

\bibitem{brose2008janimalecology}Brose, U., Ehnes, R.B., Rall, B.C., Vucic‐Pestic, O., Berlow, E.L. and Scheu, S., Journal of Animal Ecology. \textbf{77(5)}(2008)

\bibitem{choudhary2015epl}Choudhary, A., Venkataraman, D., Ray, S. S., EPL (Europhysics Letters), \textbf{112(2)} (2015)

\bibitem{Scholzetal2018natcomm}Scholz, C., Jahanshahi, S., Ldov, A., Löwen, H, Nature communications, {\bf 9(1)} (2018)

\bibitem{cavagnaetal2015jsp}Cavagna, A., Del Castello, L., Giardina, I., Grigera, T., Jelic, A., Melillo, S., ... Walczak, A. M,  Journal of Statistical Physics. \textbf{158(3)} (2015)

\bibitem{ZhdankinPRE2010}Zhdankin V, Sprott JC, Physical Review E. 82(5), (2010)

\bibitem{milanjanosov2017njp}Janosov, M., Virágh, C., Vásárhelyi, G. and Vicsek, T., New Journal of Physics. {\bf 19(5)} (2017)

\bibitem{kerleyjoz2005}Hayward MW, Kerley GI, Journal of zoology. \textbf{267(3)} (2005)

\bibitem{Mckenzieinterfacefocus2012}McKenzie HW, Merrill EH, Spiteri RJ, Lewis MA, Interface focus. \textbf{2(2)}(2012)

\bibitem{carobook2005}Caro T. Antipredator defenses in birds and mammals. University of Chicago Press(2005).

\bibitem{Humpherisoecologia1970}Humphries DA, Driver PM, Oecologia. \textbf{5(4)} (1970)

\bibitem{DomeniciJEB2011}Domenici P, Blagburn JM, Bacon JP, Journal of Experimental Biology. \textbf{214(15)}(2011)

\bibitem{EdutBBR2004}Edut S, Eilam D. Behavioural brain research. {\bf 155(2)} (2004).




\bibitem{DomeniciMB1997}Domenici P, Batty RS, Marine Biology. \textbf{128(1)} (1997).

\bibitem{pavlov2000JOI}Pavlov, D. S., Kasumyan, A. O,  Journal of Ichthyology, \textbf{40(2)}(2000).

\end{thebibliography}
\end{document}